\documentclass[journal,12pt,onecolumn,draftclsnofoot,]{IEEEtran}
\usepackage[utf8]{inputenc}
\usepackage{amsmath}
\usepackage{dsfont}
\usepackage{bbm}
\usepackage{amsfonts}
\usepackage{algorithm}
\usepackage{physics}
\usepackage{algpseudocode,graphicx}
\usepackage{amsthm}
\usepackage{mathtools}
\usepackage{hyperref}
\usepackage{xcolor}
\usepackage{dsfont}
\usepackage{caption}
\usepackage{subcaption,etoolbox}
\usepackage{cite}

\newtheorem{assumption}{Assumption}

\newtheorem{theorem}{Theorem}

\usepackage{amsthm}
\usepackage{stmaryrd,scalerel} 
\usepackage{enumitem}

\setlist[itemize]{leftmargin=*}
\DeclareMathOperator*{\minimize}{minimize}
\DeclareMathOperator*{\maximize}{maximize}
\DeclarePairedDelimiter\ceil{\lceil}{\rceil}

\title{Forking Uncertainties: Reliable Prediction and Model Predictive Control with Sequence Models via  Conformal Risk Control}
\author{Matteo Zecchin, Sangwoo Park, and Osvaldo Simeone
\thanks{
Matteo Zecchin, Sangwoo Park, and Osvaldo Simeone are with the  King’s Communications, Learning \& Information Processing (KCLIP) lab within the Centre for Intelligent Information Processing Systems (CIIPS), Department of Engineering, King’s College London, London WC2R 2LS, U.K. (e-mail: matteo.1.zecchin@kcl.ac.uk; sangwoo.park@kcl.ac.uk; osvaldo.simeone@kcl.ac.uk). 
(sho
The work of M. Zecchin and O. Simeone was supported by the European Union’s Horizon Europe project CENTRIC (101096379). The work of O. Simeone was also supported by the Open Fellowships of the EPSRC (EP/W024101/1) by the EPSRC project  (EP/X011852/1), and by Project REASON, a UK Government funded project under the Future Open Networks Research Challenge (FONRC) sponsored by the Department of Science Innovation and Technology (DSIT).

}}

\begin{document}
\maketitle

\begin{abstract}
    In many real-world problems, predictions are leveraged to monitor and control cyber-physical systems, demanding guarantees on the satisfaction of reliability and safety requirements. However, predictions are inherently uncertain, and managing prediction uncertainty presents significant challenges in environments characterized by complex dynamics and forking trajectories. In this work, we assume access to a pre-designed probabilistic implicit or explicit sequence model, which may have been obtained using model-based or model-free methods. We introduce probabilistic time series-conformal risk prediction (PTS-CRC), a novel post-hoc calibration procedure that operates on the predictions produced by any pre-designed  probabilistic forecaster to yield reliable error bars.  In contrast to existing art,  PTS-CRC produces predictive sets based on an ensemble of multiple prototype trajectories sampled from the sequence model, supporting the efficient representation of forking uncertainties. Furthermore, unlike the state of the art, PTS-CRC can satisfy reliability definitions beyond coverage. This property is leveraged to devise a novel model predictive control (MPC) framework that addresses open-loop and closed-loop control problems under general average constraints on the quality or safety of the control policy. We experimentally validate the performance of PTS-CRC prediction and control by studying a number of use cases in the context of wireless networking. Across all the considered tasks, PTS-CRC predictors are shown to provide more informative predictive sets, as well as safe control policies with larger returns.
\end{abstract}

\section{Introduction}
\label{sec:intro}
\subsection{Motivation and Overview}
\begin{figure}[t]
     \centering
     \begin{subfigure}[b]{0.32\textwidth}
         \centering
         \includegraphics[width=\textwidth]{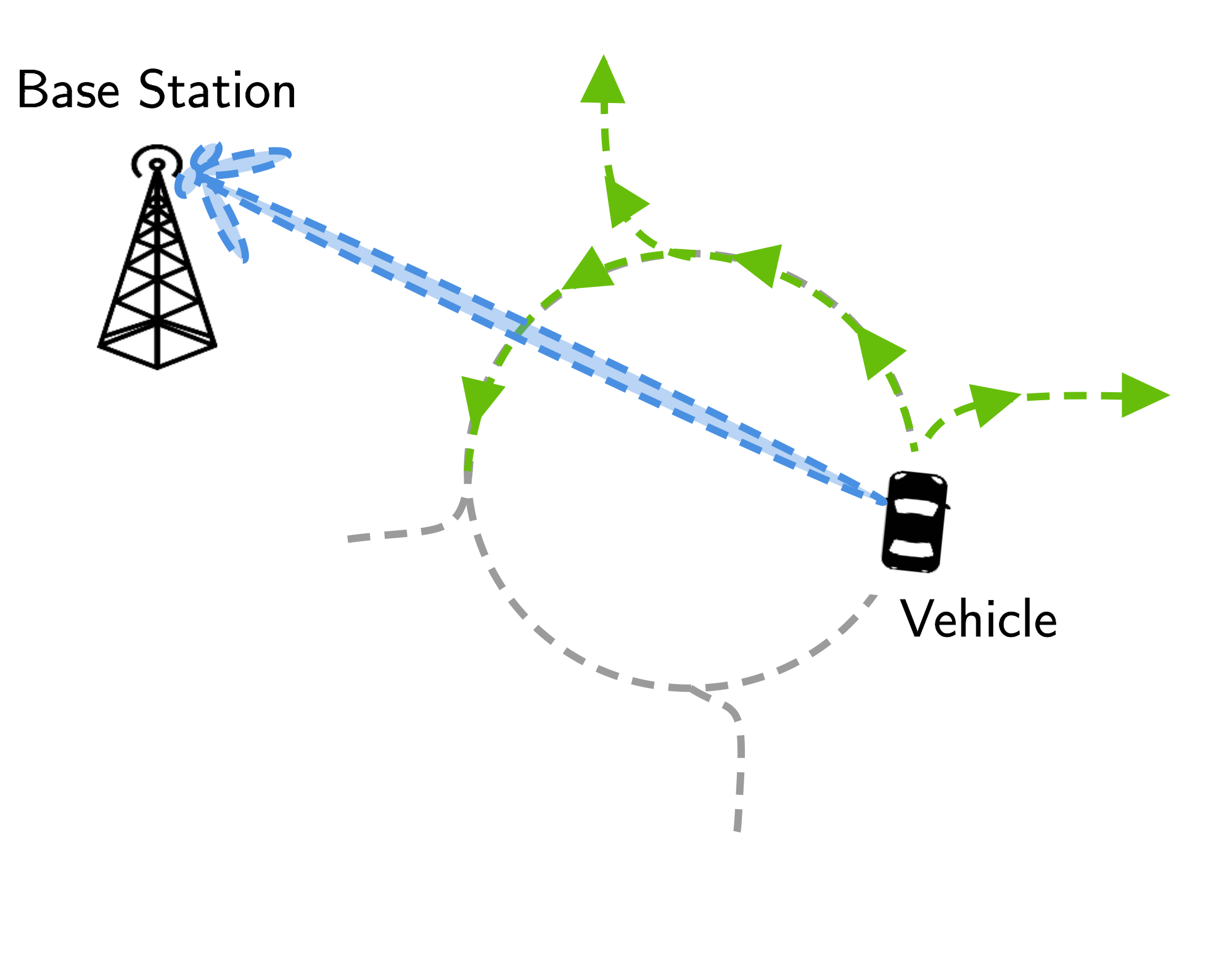}
         \caption{Scenario\label{fig:scenario}}
     \end{subfigure}
     \hfill
     \begin{subfigure}[b]{0.32\textwidth}
         \centering
         \includegraphics[width=\textwidth]{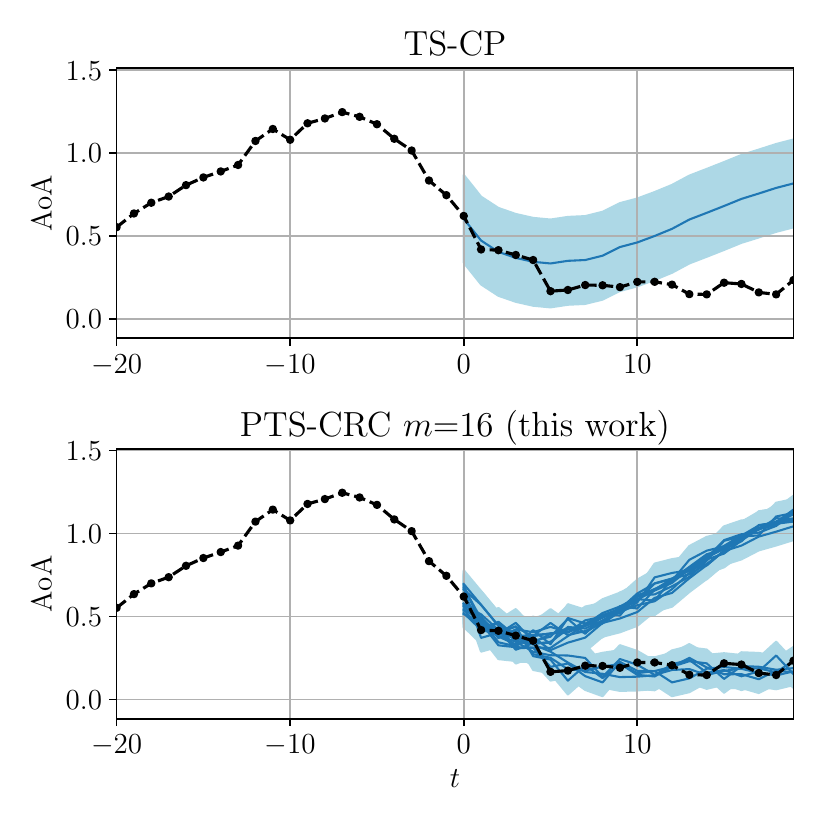}
         \caption{Prediction example\label{fig:prediction_example}}
     \end{subfigure}
     \hfill
     \begin{subfigure}[b]{0.32\textwidth}
         \centering
         \includegraphics[width=\textwidth]{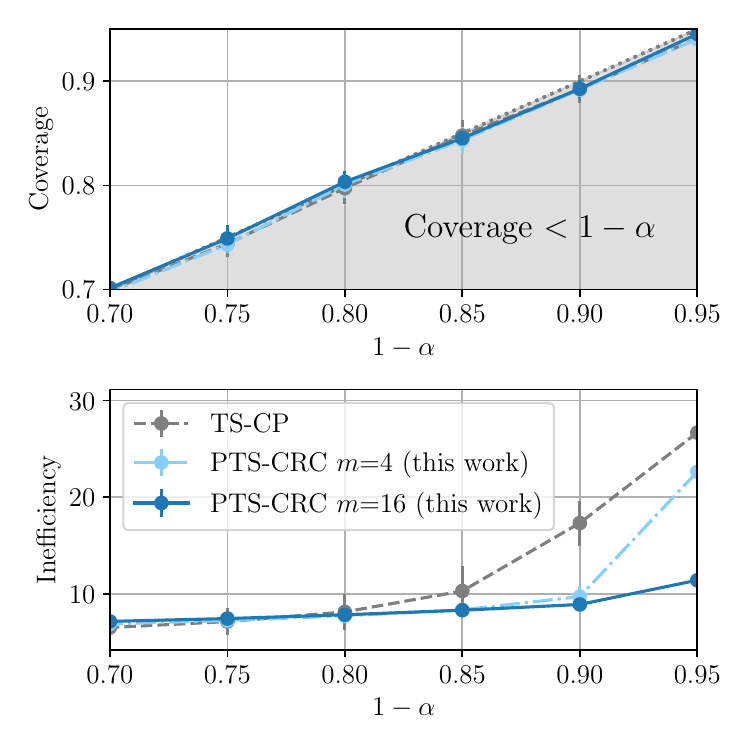}
         \caption{Performance\label{fig:performance}}
     \end{subfigure}
    \caption{Illustration of an exemplifying application of the proposed method: (a) The base station wishes to predict the evolution of the angle of arrival $y_t$ for the line-of-sight radio propagation path from a vehicle moving in a roundabout. (b) Given knowledge of the sequence $y_t$ from at times $t=-20,\dots,-1$,  the goal is to produce prediction intervals for the future evolution of the sequence $y_t$ at times $t=0,...,19$. The true trajectory is shown as a dashed black line, with the  predictive intervals (blue shaded areas) produced by the state-of-the-art TS-CP \cite{stankeviciute2021conformal} provided in the upper part,  while  the proposed set-predictor (PTS-CRC) with $m=4$ and $m=16$ prototypes (see Fig. \ref{fig:prototype_set_predictor}) are plotted in the lower part.  (c) Both TS-CP and PTS-CRC are guaranteed to cover the true trajectory with probability at least $1-\alpha$ (top), but PTS-CRC significantly reduces the inefficiency, i.e., the average predicted set size  (bottom).  }
       \label{fig:AoA} 
\end{figure}

 In many real-world problems, predictions are leveraged to monitor and control cyber-physical systems. For instance, when planning the path of a robot on the floor of a factory, one may leverage   predictions of the other robots' or agents' movements   \cite{ji2016path,wang2021risk}. As another example, designing the trajectory of a drone  for the purpose of data collection, tracking, or providing wireless connectivity may benefit from  access to predictions regarding future data generation, target movements, or wireless traffic levels \cite{lee2003strategies,vanegas2017uav,li2020path}. And, as illustrated in Fig. \ref{fig:AoA}, in wireless networks, in order to point a transmission or reception  beam towards a user at high carrier frequencies, a base station can make use of the predicted angle of departure or arrival of the wireless signal \cite{moon2020deep,lim2021deep}.

Predictions are inherently uncertain, and yet the  monitoring and  operation of cyber-physical systems typically demand guarantees on the satisfaction of reliability and safety requirements. 
With reference to the examples above, in the context of path planning of autonomous agents, robots should be able to operate while avoiding collision and respecting satisfy safety margins \cite{kunchev2006path}. Drones have to ensure a minimal level of coverage in mission critical scenarios  such as natural disasters, e.g., for surveillance or aid delivery \cite{patterson2014timely,zeng2016wireless}. In 5G wireless  systems, base stations have to reliably forecast traffic demands and channel state evolutions to support ultra-reliable low latency communications (URLLC) applications \cite{chen2018ultra}.

While essential for the reliable monitoring and safe control of cyber-physical systems, managing prediction uncertainty presents  significant challenges in environments characterized by complex dynamics. For instance, in the setting  illustrated in Fig. \ref{fig:AoA}, a vehicle may exit a roundabout at several intersections, creating forks in the possible future trajectories, with each possible trajectory having error bars of its own,  accounting, e.g., for variable accelerations or decelerations. 

Existing approaches for the design of predictors and of \emph{model predictive control} (MPC) systems encompass \emph{model-based} and \emph{model-free} methods. In the former case,  domain knowledge is leveraged to define a physically motivated model of the environment, from which  predictions and optimal policies are constructed. Any reliability or safety guarantee  provided by this class of techniques is valid only as long as the underlying system's model is accurate \cite{brosilow2002techniques,rossiter2017model}.  Model-free solutions offer an alternative approach, whereby  environment dynamics are captured by using general-purpose models -- such as autoregressive models or transfomer-based solutions like large language models (LLMs) -- that are optimized based on data. The higher versatility of model-free methods comes at the cost of weaker guarantees and costly data collection procedures \cite{garcia2015comprehensive,hasanbeig2020towards}. 

In this work, we assume access to a \emph{pre-designed probabilistic sequence model}, which may have been obtained by using model-based or model-free methods. As an example, one may have trained a transformer-based sequence model such as an LLM or a decision-based transformer \cite{lim2021temporal,tang2021probabilistic,chen2021decision}, or one may have optimized a Kalman predictor, or variants thereof, based on a dynamic model \cite{ristic2003beyond}. As such, the proposed approach takes an agnostic stance toward modeling assumptions, only positing the availability of any \emph{implicit } or \emph{explicit} probabilistic model. An implicit model can only generate predictive samples, as is the case for models based on complex physics simulations \cite{cranmer2020frontier} and for neural flow algorithms \cite{kingma2016improved}; while an explicit model can also assign probabilities to the possible predictive outputs, as for Kalman filters or language models. We propose a novel general framework, grounded in \emph{conformal risk control} (CRC) \cite{angelopoulos2022conformal}, that enables (\emph{i}) efficient and provably reliable predictive uncertainty quantification even in situations with forking paths (see Fig. \ref{fig:AoA}); and (\emph{ii}) MPC with guarantees on general average constraints with respect to the future behavior of the system.  To validate the theoretical properties of the proposed methodology, we showcase several applications related to the monitoring and control of modern wireless systems.

\subsection{Related Work}
Quantifying uncertainty in the prediction of time series and ensuring the safe operation of complex systems are fundamental problems in statistics and control theory, respectively. Traditional \emph{model-based} approaches are based on autoregressive modeling assumptions, including Kalman filters, and on frameworks such as control Lyapunov theory, control barrier functions, and robust MPC \cite{romdlony2016stabilization,anand2021safe,saltik2018outlook}. All these strategies  offer theoretical guarantees under the assumptions that the postulated models are valid \cite{bemporad2007robust} and that the computational complexity of the optimized strategies, particularly in the case of non-linear system dynamics, affords an efficient implementation  \cite{manchester2018robust}. 

Data-driven control approaches, including recurrent neural networks (RNNs) \cite{rumelhart1985learning}, long short-term memory networks (LSTMs) \cite{hochreiter1997long} and LLMs  \cite{vaswani2017attention}, learning-based MPC \cite{hewing2020learning}, and model-free reinforcement learning (RL) \cite{chen2023probabilistic,hasanbeig2020towards,garcia2015comprehensive}, can potentially address these shortcomings and have gained significant attention over the past decades. Learning-based methods exhibit remarkable versatility, but they provide looser guarantees as compared to their model-based counterparts, being ultimately limited by the availability of relevant data. A particular concern in this regard is that machine learning models  may excel at accurate prediction of complex dynamics under best-case conditions, but they typically lack reliable \emph{uncertainty quantification} capabilities \cite{guo2017calibration,tao2023benchmark}. Bayesian variants of these forecasting models offer a principled approach to account for uncertainty. However, they are limited by the approximations required for efficient implementations \cite{simeone2022machine} and they are susceptible to model misspecification and outliers \cite{walker2013bayesian,martinez2018practical,zecchin2023robust}. 

In this context, \emph{conformal prediction} (CP) has recently emerged as a prominent \emph{post-hoc calibration} technique. CP can be applied to the output of any pre-designed model to yield reliable error bars with (frequentist) finite-sample coverage guarantees \cite{vovk2005algorithmic,quach2023conformal,deutschmann2023conformal}. In its most practical incarnation, CP requires the use of a separate calibration data set \cite{angelopoulos2022conformal}, and it may leverage probabilistic predictors \cite{wang2022probabilistic}. CP has been recently applied to the problem of quantifying predictive uncertainty for  time series  in various domains. These include the  monitoring  of LLM-based planning routines \cite{wang2023conformal,ren2023robots}, the estimate of the returns of policies for Markov decision processes \cite{dietterich2022conformal}, and the run-time verification of dynamic systems  \cite{lindemann2023conformal,cairoli2023conformal}. The error bars produced by CP have also been leveraged for MPC targeting  the safe planning of agents in shared environments \cite{lindemann2022safe,dixit2023adaptive}.

\subsection{Contributions}

In this work, we address the problems of reliable prediction and safe MPC by leveraging pre-designed implicit or explicit probabilistic sequence models and calibration data in the form of sample time series for the quantities  being predicted. Reliable prediction requires the evaluation of error bars satisfying average accuracy constraints, while safe MPC entails the satisfaction of average risk constraints. As summarized in the previous subsection, existing art proposed the  application of CP to  tackle these challenges, resulting in the following two limitations that we aim to address in this work. 
\begin{itemize}
\item \emph{Unimodal vs. forking uncertainties}: In the example in Fig. \ref{fig:AoA}, the base station wishes to predict the evolution of the angle of arrival 
 for the line-of-sight radio propagation path between a moving vehicle in a roundabout and the base station. Given that the vehicle can exit the roundabout at any of the four side streets, the future evolution of the angle of arrival has multiple  possible forking trajectories. State-of-the-art  \emph{time series-CP} (TS-CP) \cite{stankeviciute2021conformal} builds a unimodal error bar around a single trajectory (Fig. \ref{fig:prediction_example},  upper part), and it hence cannot capture the \emph{forking uncertainties} caused by the multiple possible future evolutions of the target process. This causes the predictive interval to be inefficient, i.e., excessively large, in order to provide coverage guarantees (Fig. \ref{fig:performance}).
\item \emph{Coverage vs. risk}: CP targets reliability guarantees in terms of coverage probability, i.e., of the probability that the error bars include the true future trajectory. In control applications, one may need to address more general average risk constraints that cannot be expressed in terms of coverage probability. For instance, one may wish to impose that the average quality   of a predicted text crosses a user-defined threshold or that the average quality of service provided by the base station in Fig. \ref{fig:AoA} be sufficiently large. Existing methods cannot address such constraints. 
\end{itemize}

Targeting these two limitations of the state of the art, the contributions of this work are as follows.
\begin{itemize}
    \item \emph{Probabilistic time series conformal risk control (PTS-CRC)}: We introduce PTS-CRC, a post-hoc calibration procedure that operates on the predictions produced by any pre-designed  sequence model to yield reliable error bars.  In contrast to existing art,  PTS-CRC produces predictive sets based on an ensemble of multiple prototype trajectories sampled from the sequence model (Fig. \ref{fig:prediction_example}, lower part, and Fig. \ref{fig:prototype_set_predictor}). This way, PTS-CRC can efficiently account for forking uncertainties. Furthermore, building on CRC \cite{angelopoulos2022conformal}, PTS-CRC can satisfy reliability definitions beyond coverage. 
    \item  \emph{PTS-CRC-based safe MPC}: We introduce a novel MPC framework that addresses open-loop and closed-loop control problems under general average constraints on the quality or safety of the control policy. The approach builds on the set predictors produced by PTS-CRC.
    \item \emph{Use cases}: We experimentally validate the performance of PTS-CRC prediction and MPC by studying a number of use cases in the context of wireless networking. First, we consider the problem of reliably monitoring the evolution of the channel gain between a base station and a moving user in an urban microcell scenario. Second, we leverage the proposed PTS-CRC-based control framework to design power control policies that satisfy reliability requirements, namely maximum interference constraints and minimal decoding packet probability guarantees. Across all the considered tasks, PTS-CRC predictors are shown to provide more informative predictive sets and safe control policies with larger returns.
\end{itemize}
The rest of the paper is organized as follows. In Section \ref{sec:prob_def}, we formally describe the prediction and MPC settings under study.  In Section \ref{background:tscp}, we review TS-CP \cite{stankeviciute2021conformal}. In Section \ref{sec:ptscp}, we introduce and analyze PTS-CRC for both implicit and explicit sequence models.  In Section \ref{sec:mpc}, we present novel control algorithms that leverage PTS-CRC for MCP. In Section \ref{sec:derivation}, we elucidate the relations between the proposed PTS-CRC and the frameworks of probabilistic CP \cite{wang2022probabilistic} and CRC \cite{angelopoulos2022conformal}. Section \ref{sec:experiments}  experimentally validates the proposed framework and the paper is concluded in Sec. \ref{sec:conclusions}.

\section{Problem Definition}
\label{sec:prob_def}

In this section, we describe first the setting and performance criteria for the problem of reliable prediction, and then we detail the class of MPC problems under study.

\subsection{Prediction}
\label{sec:setting}
Given the first $T$ samples of a  time series $y_{-T},...,y_{-1}$, we are interested in predicting the next $\tau$ samples $y_{0},...,y_{\tau-1}$ by providing sets, i.e., ``error bars'', that satisfy reliability guarantees with respect to the correct trajectory (see Fig. \ref{fig:AoA}). The overall trajectory $y_{-T:\tau-1}=[y_{-T},\dots,y_{\tau-1}]\in \mathcal{Y}^{T+\tau}$ is a time series in which each sample $y_t$ takes values in a set $\mathcal{Y}$. Set $\mathcal{Y}$ may be discrete and finite or a subset of a real-valued vector space.
Its distribution $p(y_{-T:\tau-1})$ is \emph{unknown}, and it can be generally written using the chain rule as 
 \begin{align}
     p(y_{-T:\tau-1})=p(y_{-T})\prod^{\tau-1}_{t=-T+1} p(y_{t}|y_{-T:t-1}),
     \label{joint}
 \end{align}
where $p(y_t|y_{-T:t-1})$ is the conditional distribution of sample $y_t$ given the past samples $y_{-T:t-1}$.

We assume access to a sequence model in the form of a \emph{probabilistic predictor} described by a conditional distribution $\hat{p}({y}_{0:\tau-1}|y_{-T:-1})$ over the future samples $y_{0:\tau-1}$ given the past samples $y_{-T:-1}$. The predictor $\hat{p}({y}_{0:\tau-1}|y_{-T:-1})$ is an approximation of the true conditional distribution $p({y}_{0:\tau-1}|y_{-T:-1})$ obtained from the joint distribution in (\ref{joint}). As mentioned in Section \ref{sec:intro}, the conditional distribution $\hat{p}({y}_{0:\tau-1}|y_{-T:-1})$ may, e.g.,  take the form of a pre-trained machine learning model or of a  model-based predictor based on a physics-driven simulator (see, e.g., \cite{jose2020address, revach2022kalmannet, pratik2021neural}).
Furthermore, as detailed next, while prior work \cite{stankeviciute2021conformal,sun2022copula,dietterich2022conformal,cleaveland2023conformal,lindemann2022safe,lindemann2023conformal} focused on deterministic predictors, here we assume either implicit or explicit probabilistic models.
\begin{itemize}
\item \emph{Deterministic predictors}: Prior work \cite{stankeviciute2021conformal,sun2022copula,dietterich2022conformal,cleaveland2023conformal,lindemann2022safe,lindemann2023conformal} assumed that the predictor $\hat{p}({y}_{0:\tau-1}|y_{-T:-1})$  is concentrated at a single predicted trajectory  $\hat{y}_{0:\tau-1}\in \mathcal{Y}^{\tau}$, which is thus a deterministic function of the past samples $y_{-T:-1}$. That is, we have the equality $\hat{p}(y_{0:\tau-1}|y_{-T:-1})=\delta\left({y}_{0:\tau-1}-\hat{y}_{0:\tau-1}\right)$,
 where $\delta(\cdot)$ is the Kronecker or Dirac delta function depending on whether the domain $\mathcal{Y}$ is discrete or continuous, respectively. 
\item \emph{Implicit, or generative-only, probabilistic predictors}: Implicit probabilistic predictors can generate predicted trajectories  \begin{equation}\label{eq:samples}\hat{y}_{0:\tau-1}\sim \hat{p}({y}_{0:\tau-1}|y_{-T:-1}),\end{equation} which are conditionally independent given the input  $y_{-T:-1}$. Such models do not provide an explicit value for the conditional distribution $\hat{p}(\hat{y}_{0:\tau-1}|y_{-T:-1})$ for the generated samples $\hat{y}_{0:\tau-1}$. Examples include time-GANs \cite{yoon2019time}, diffusion models \cite{rasul2021autoregressive}, and probabilistic spiking neural networks \cite{jang2019introduction,rosenfeld2022spiking}.
\item \emph{Explicit, or likelihood-based, probabilistic predictors}: Explicit probabilistic predictors can generate samples (\ref{eq:samples}) like implicit models, but they also provide as output the value $ \hat{p}(\hat{y}_{0:\tau-1}|y_{-T:-1})$ assigned by the model to the generated sample $\hat{y}_{0:\tau-1}$.
\end{itemize}

\subsection{Set Prediction}

The goal of this work is to leverage the available, implicit or explicit, probabilistic predictor $\hat{p}({y}_{0:\tau-1}|y_{-T:-1})$ to produce a prediction set over the space of future sequences. We focus on the special class of set predictors that depend on the past evolution $y_{-T:-1}$ through a set $\mathcal{P}^m(y_{-T:-1})=\{{y}^i_{0:\tau-1}\}^m_{i=1}$ of $m$ \emph{prototypical sequences}, where $y_{0:\tau-1}^i$ is the $i$-th prototype.  While more general forms of the predictors are possible, we will consider set predictors that include all future sequences $y_{0:\tau-1}$ that are sufficiently close to any of the prototypes  with respect to a given distance measure $d(\cdot,\cdot)$, i.e., 
\begin{subequations}
\begin{align}
    \Gamma(y_{-T:-1})&=\left\{y_{0:\tau-1}\in\mathcal{Y}^\tau: \min_{\hat{y}_{0:\tau-1}\in \mathcal{P}^m(y_{-T:-1})}d(y_{0:\tau-1},\hat{y}_{0:\tau-1})\leq \lambda\right\} \label{3a}\\
    &=\bigcup_{\hat{y}_{0:\tau-1}\in \mathcal{P}^m(y_{-T:-1})}\left\{y_{0:\tau-1}\in\mathcal{Y}^\tau: d(y_{0:\tau-1},\hat{y}_{0:\tau-1})\leq \lambda\right\},
    \label{3b}
\end{align}
    \label{eq:example_proto_setsequencepredictor}
\end{subequations}
where $\lambda>0$ is a design parameter. The equality between (\ref{3a}) and (\ref{3b}) follows from the fact that, if a sequence $y_{0:\tau-1}$ is in the set defined by (\ref{3a}), there exists a prototype $\hat{y}_{0:\tau-1}\in\mathcal{P}^m(y_{-T:-1})$ satisfying the inequality $d(y_{0:\tau-1},\hat{y}_{0:\tau-1})\leq \lambda$,  ensuring that the sequence is also in set (\ref{3b}), and vice versa. As we will discuss in Section \ref{sec:ptscp}, existing works \cite{stankeviciute2021conformal,sun2022copula,dietterich2022conformal,cleaveland2023conformal,lindemann2022safe,lindemann2023conformal} only consider the case $m=1$, while this paper leverages the use of probabilistic predictors to allow for $m>1$ prototypes.

\begin{figure}[t]
     \centering
         \centering
         \includegraphics[width=0.6\textwidth]{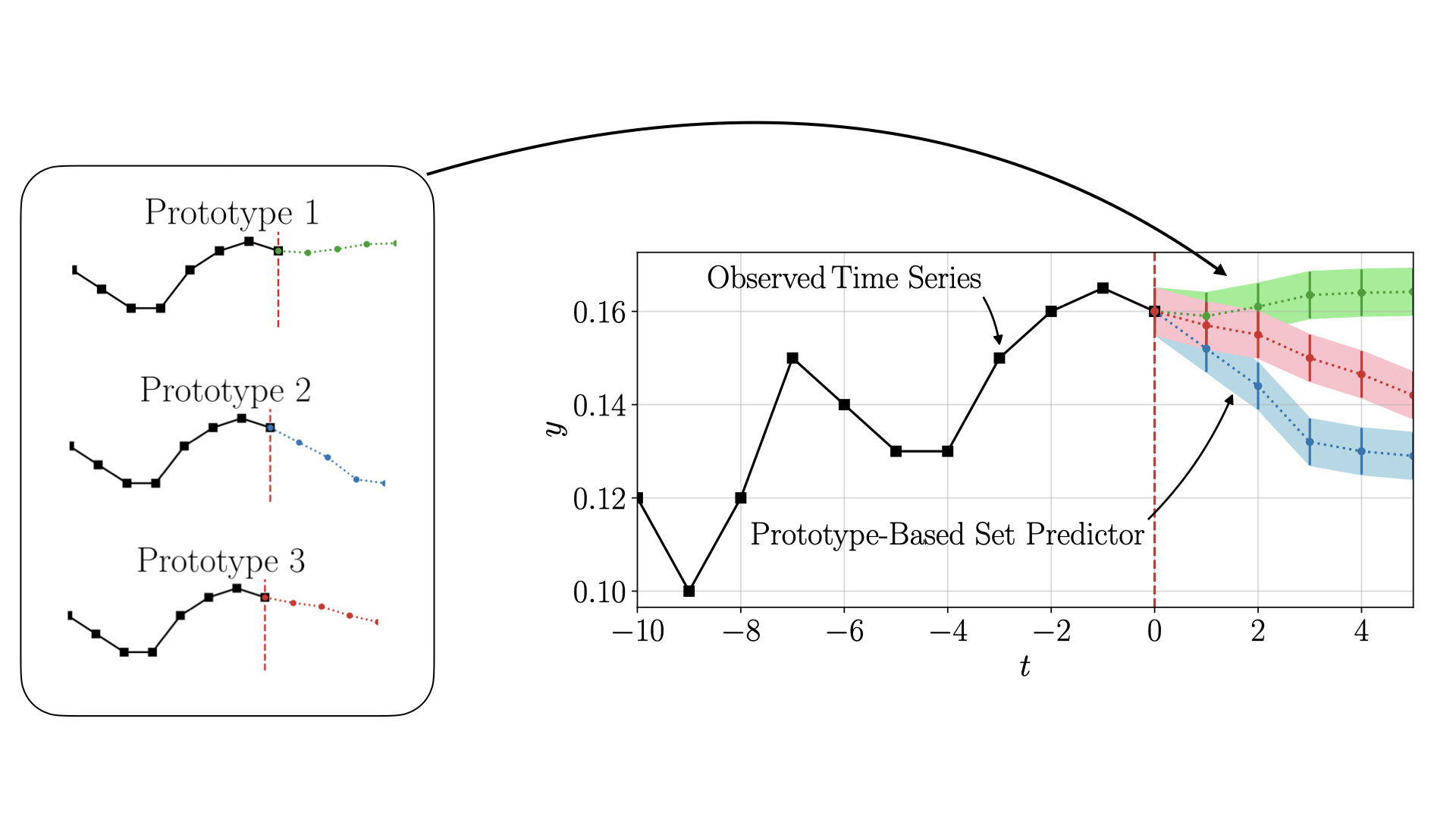}
    \caption{A prototype-based set predictor: Given the past evolution of time series $y_{-T:-1}$ (solid black curve), the prototype-based set predictor $\Gamma(y_{-T:-1})$ in (\ref{eq:example_proto_setsequencepredictor}) is constructed based on the set $\mathcal{P}^m(y_{-T:-1})$ containing $m$ prototypical sequences (here $m=3$) by including all sequences $\hat{y}_{0:\tau-1}$ whose maximum distance to one of the prototypes is bounded by $\lambda$ (here $\lambda=0.05$). The prototype-based set predictor $\Gamma(y_{-T:-1})$, reported on the right, includes all the sequences that are either in the red, green, or blue-shaded areas. While prior work is limited to the case of a single prototype  ($m=1$), this paper leverages the use of probabilistic sequence models to allow for $m>1$ prototypes.}
        \label{fig:prototype_set_predictor}
\end{figure}

In Fig. \ref{fig:prototype_set_predictor} we provide an illustration of a prototype-based set predictor $\Gamma(y_{-T:-1})$ of the form (\ref{eq:example_proto_setsequencepredictor}) obtained from a set of $m=3$ prototypical sequences.
The set predictor expresses the predictor's expectation that the future sequence $y_{0:\tau-1}$ is in set $\Gamma(y_{-T:-1})$ given the available information $y_{-T:-1}$ as input. As per (\ref{eq:example_proto_setsequencepredictor}), the set predictor $\Gamma(y_{-T:-1})$ can be obtained by including all sequences $y_{0:\tau-1}$ whose distance from any prototype $\hat{y}_{0:\tau-1}$ is no larger than $\lambda$.

Given a set predictor $\Gamma(y_{-T:-1})$ we define the \emph{per-time step predicted set} at time $t$ as including all values of $\mathcal{Y}$ that are assumed by some trajectory $\hat{y}_{0:\tau-1}$ in set $\Gamma(y_{-T:-1})$ at time $t$, i.e.,
\begin{align}
    \Gamma_t(y_{-T:-1})=\left\{y\in\mathcal{Y}: \exists \ \hat{y}_{0:\tau-1}\in \Gamma(y_{-T:-1}) \text{ with } \hat{y}_{t}=y\right\}.
    \label{eq:setsequencepredictor_at_time_t}
\end{align}

\subsection{Reliability}
We are interested in producing interval predictors $\Gamma(y_{-T:-1})$ that are \emph{reliable} and \emph{efficient}. To define \emph{reliability}, let us fix a \emph{loss function}   $\mathcal{L}\left(\Gamma,y_{0:\tau-1}\right) $ that measures the discrepancy between the predicted set $\Gamma$ and the actual trajectory realization $y_{0:\tau-1}$. We impose some mild assumptions on the loss function in a manner similar to \cite{angelopoulos2022conformal}.
\begin{assumption}[Bounded and monotonic loss function]
    \label{ass:loss}
    The loss function  $\mathcal{L}\left(\cdot,\cdot\right)$ satisfies the inequality 
    \begin{align}
    \mathcal{L}\left(\Gamma,y_{0:\tau-1}\right)\leq B
    \label{eq:boundedloss}
    \end{align} 
    for all pairs $(\Gamma,y_{0:\tau-1})$, where $B\leq \infty$ is a constant, and it is monotonic in the size of the predicted set, i.e., we have the inequality $\mathcal{L}\left(\Gamma,y\right)\leq \mathcal{L}\left(\Gamma',y\right)$ for any pair of sets satisfying the inclusion relation $\Gamma'\subseteq\Gamma$.
\end{assumption}

We say that the set predictor $\Gamma(\cdot)$ is $\alpha$-\emph{reliable} with respect to the loss $\mathcal{L}(\cdot,\cdot)$ if the expected value of the loss, $\mathcal{R}(\Gamma)$,  is bounded by a \emph{target maximum unreliability level}  $\alpha$. Accordingly, the predictor $\Gamma(y_{-T:-1})$ is $\alpha$-reliable if it satisfies inequality
\begin{align}
    \mathcal{R}(\Gamma):=\mathbb{E}[\mathcal{L}\left(\Gamma(y_{-T:-1}),y_{0:\tau-1}\right)]\leq \alpha,
    \label{eq:reliabilityguarantee}
\end{align}
where the expectation is over the unknown distribution $p(y_{-T:\tau-1})$ in (\ref{joint}) of the time series $y_{-T:\tau-1}$.

The definition of reliability (\ref{eq:reliabilityguarantee}) specializes to distinct requirements depending on the choice of the loss functions. Some important examples, satisfying Assumption \ref{ass:loss}, are as follows.
\begin{itemize}
    \item \emph{Sequence coverage probability}: The \emph{miscoverage} loss function
    \begin{align}
    \label{eq:miscoverageloss}
\mathcal{L}\left(\Gamma(y_{-T:-1}),y_{0:\tau-1}\right)=\mathds{1}\left\{y_{0:\tau-1} \notin \Gamma(y_{-T:-1}) \right\}
    \end{align} returns 1 if the actual sequence $y_{0:\tau-1}$ of future samples is not included in the predicted set $\Gamma(y_{-T:-1})$. With this loss function, characterized by $B=1$ in 
    (\ref{eq:boundedloss}), the reliability requirement (\ref{eq:reliabilityguarantee}) corresponds  to the \emph{sequence coverage probability} guarantee \cite{stankeviciute2021conformal,sun2022copula,dietterich2022conformal,cleaveland2023conformal,lindemann2022safe,lindemann2023conformal}
    \begin{align}
       \Pr[y_{0:\tau-1}\notin \Gamma(y_{-T:-1})]\leq\alpha.
       \label{eq:coverage_guarantee}
    \end{align}
    \item \emph{Sample coverage rate}: The \emph{per-sample miscoverage rate} loss function
        \begin{align}\label{eq:persampleloss}
            \mathcal{L}\left(\Gamma(y_{-T:-1}),y_{0:\tau-1}\right)=\frac{1}{\tau}\sum^{\tau-1}_{t=0}\mathds{1}\left\{y_{t} \notin \Gamma_t(y_{-T:-1}) \right\}
        \end{align} calculates the fraction of future samples $\{y_{t}\}_{t=0}^{\tau-1}$ that are not included in the corresponding per-time step predicted subsets $\{\Gamma_t(y_{-T:-1})\}_{t=0}^{\tau-1}$ in (\ref{eq:setsequencepredictor_at_time_t}). With loss (\ref{eq:persampleloss}), also characterized by $B=1$ in 
    (\ref{eq:boundedloss}),  the reliability condition (\ref{eq:reliabilityguarantee}) reduces to the constraint
        \begin{align}
           \frac{1}{\tau}\sum^{\tau-1}_{t=0}\Pr\left[y_{t} \notin \Gamma_t(y_{-T:-1}) \right]\leq \alpha
           \label{ta_coverage_guarantee}
        \end{align}
        that, on average, the  prediction sets contain the future trajectory for at least a fraction $1-\alpha$ of the future time steps. Note that the requirement (\ref{ta_coverage_guarantee}) is less strict than (\ref{eq:coverage_guarantee}).
\end{itemize}

\subsection{Efficiency}
Efficiency refers to the informativeness of the interval prediction $\Gamma(\cdot,\cdot)$  which is generally measured by its size \cite{stutz2021learning, dhillon2023expected}. To appreciate the tension between reliability and efficiency, note that the coverage guarantee (\ref{eq:coverage_guarantee}) can be satisfied for any value of $\alpha\in[0,1]$ by the trivial interval  predictor that always outputs the entire space $\Gamma(y_{-T:-1})=\mathcal{Y}^\tau$ of possible trajectories. This prediction is clearly not informative, but it is perfectly reliable.

 To formalize the notion of inefficiency, we fix a measure $\mu(\cdot)$ over the space $\mathcal{Y}^\tau$ of trajectories. For instance, if set $\mathcal{Y}$ is discrete, measure $\mu(\Gamma)$ may count the number of trajectories in subset $\Gamma\subseteq \mathcal{Y}^\tau$; or if $\mathcal{Y}^\tau=\mathbb{R}^\tau$, function $\mu(\Gamma)$ may be the Lebesgue measure or the time-averaged measure $(1/\tau)\sum^{\tau-1}_{t=0} \mu(\Gamma_t)$ evaluated using the per-time step predictions (\ref{eq:setsequencepredictor_at_time_t}). 
The \emph{inefficiency} of a set predictor  $\Gamma(\cdot)$ is then defined as the average size 
\begin{align}
   \mathcal{I}(\Gamma):=\mathbb{E}[\mu\left(\Gamma(y_{-T:-1})\right)],
   \label{inefficiency}
\end{align} where the average is again over the unknown distribution $p(y_{-T:\tau-1})$ in (\ref{joint}).

\subsection{Model Predictive Control}
\label{sec:mpc_intro}

Reliable and efficient set predictors can support decision making processes that are subject to reliability or safety requirements. In this work, we specifically focus on the general problem of controlling a dynamical system whose \emph{state} is described by a  state variable $s_t \in \mathcal{S}$ via the selection of a sequence of \emph{actions} $u_t \in \mathcal{U}$ over time $t$. For instance, the state $s_t$ may represent the trajectory of a drone or robot, or the occupancy of queues in a telecommunications network; while action $u_t$ may describe a steering decision or the allocation of some resources.

The evolution of the state $s_t$ depends on the action $u_t$ for time step $t=0,1,\dots$ through the system equation
\begin{align}
    s_t=f(s_{t-1},u_t)
    \label{dynamics}
\end{align} 
for some known function $f(\cdot,\cdot)$ and known initial state $s_{-1}$. The \emph{reliability}, or safety, of the sequence of states $s_t$ is measured in relation to another process $y_t$ for $t=0,1,...,\tau-1$, which we refer to as the \emph{target process}. To this end, we define a \emph{constraint function} $c(s_{0:\tau-1},y_{0:\tau-1})$, which measures the extent to which the sequence of states $s_{0:\tau-1}$ fails to meet a reliability requirement with respect to sequence $y_{0:\tau-1}$. 

For instance, for any time $t$, the value $y_t$ may represent the current distribution of data to be collected, the position of targets to be tracked, the distribution of wireless traffic levels, or the positions of other robots or agents in a shared environment. In these examples, the constraint function  $c(s_{0:\tau-1},y_{0:\tau-1})$ can measure  the extent to which the current position $s_t$ of a drone fails to cover the areas of high data availability or wireless traffic, as described by the sample $y_t$; or the distance between a robot's position $s_t$ and the positions $y_t$ of the other robots or agents.

 The target process $y_{0:\tau-1}$ has an unknown distribution (\ref{joint}). Furthermore, at any time $t$, conditioned on all past actions $u_{0:t-1}$, states $s_{0:t-1}$ and samples $y_{0:t-1}$, the next sample $y_t$ depends only on the past target samples $y_{0:t-1}$. 
That is, we have the conditional distribution 
\begin{align}
    p(y_t|y_{0:t-1},u_{0:t-1},s_{0:t-1})=p(y_t|y_{0:t-1}).
    \label{ind_proc}
\end{align}
By (\ref{ind_proc}), as also assumed in \cite{lindemann2022safe}, the target process $y_t$ is not affected by the actions $u_t$, and it is generally subject to the randomness modeled by the distribution $p(y_t|y_{-T:-t})$. 
 
The target process $y_{0:\tau-1}$ is not available to the decision maker, which has access to a sequence of past samples $y_{-T:-1}$ and to a probabilistic predictor $\hat{p}(y_{0:\tau-1}|y_{-T:-1})$, and it is aware of the initial state $s_0$. As we detail below, for closed-loop control, we also assume that, at time $t$, the controller has access to the past samples $y_{-T:\tau-1}$. Finally, note that, due to the deterministic dynamic described by (\ref{dynamics}), the controller can perfectly predict the impact of its actions $u_t$ on the evolution of the state $s_t$.

On the basis of the available information about the past samples $y_{-T:-1}$, about the initial value $s_0$, and, possibly, also about the past values $y_{0:t-1}$, the decision maker chooses each action $u_t$ for $t=0,1,\dots,\tau-1$ with the goal of minimizing the cumulative value of a cost function $J(s_t,u_t)$ over the future $\tau$ steps, while satisfying an average reliability constraint. Specifically, the decision maker addresses the problem 
\begin{subequations}
\begin{align}
\minimize_{u_{0}\dots,u_{\tau-1}} \quad & \sum^{\tau-1}_{t=0} J(s_t,u_t)\\
\textrm{s.t.} \quad & s_t=f(s_{t-1},u_t) \ \text{ for } t=0,1,\dots,\tau-1\\
&\mathbb{E}[c(s_{0:\tau-1},y_{0:\tau-1})]\leq \delta, \label{mpc_constraint}   
\end{align}
\label{mpc}%
\end{subequations} where parameters $\delta>0$ define the  \emph{maximum control unreliability level}. The constraint (\ref{mpc_constraint}) requires that the average value of the constraint function $c(s_{0:\tau-1},y_{0:\tau-1})$ to be no larger than level $\delta$, where the expectation is taken over the unknown distribution $p(y_{-T:\tau-1})$ of the target process in (\ref{joint}). 
Based on the fact that the value $\delta$ can be included in the constraint function $c(s_{0:\tau-1},y_{0:\tau-1})$, in the following we set $\delta=0$ without loss of generality.

Problem (\ref{mpc}) is addressed via MPC by choosing control actions $u_t$ that depend on predictions obtained based on the sequence model $\hat{p}(y_{0:\tau-1}|y_{-T:-1})$. Specifically, we consider both open-loop and closed-loop control formulations of problem (\ref{mpc}). The \emph{open-loop} formulation corresponds to the scenario in which the control sequence $u_{0:\tau-1}$ is designed entirely based on the information available at time $t=0$.  That is, the control sequence is evaluated by tackling problem (\ref{mpc}) based on the initial state $s_0$, the past samples $y_{-T:-1}$ and the probabilistic predictor $\hat{p}(y_{0:\tau-1}|y_{-T:-1})$. 

In the \emph{closed-loop} formulation, after every time step $t$, the controller observes the realization of $y_t$, and it can use this additional information to refine its prediction about the future evolution of the target process $y_{t:\tau-1}$. Formally, the action $u_t$ at time $t$ is given by addressing (\ref{mpc}) based on the observed state and target process evolutions up to time $t$, i.e. $s_0,\dots,s_{t-1}$ and $y_{-T},\dots,y_{t-1}$ and the predictive distribution  $\hat{p}(y_{t:\tau-1}|y_{-T:t-1})$.

Importantly, for both the open-loop and closed-loop formulations, predictions must be used  to ensure that constraint (\ref{mpc_constraint}) is satisfied with respect to the unknown target process distribution, and not with respect to the predictive distribution.

\section{Background: Time Series Conformal Prediction}
\label{background:tscp}
 In this section, we review TS-CP \cite{stankeviciute2021conformal,lindemann2023conformal,dietterich2022conformal,lindemann2022safe,cleaveland2023conformal}. TS-CP applies CP to turn a \emph{deterministic} time series forecaster into a set predictor (\ref{eq:example_proto_setsequencepredictor}) that includes the future evolution of the system with a user-specified coverage level $1-\alpha$ as per the sequence coverage probability guarantee (\ref{eq:coverage_guarantee})\cite{stankeviciute2021conformal}.

Given a predicted sequence $\hat{y}_{0:\tau-1}$, TS-CP adopts the set predictor (\ref{eq:example_proto_setsequencepredictor}) with $m=1$, where the set $\mathcal{P}^1$ of prototypes includes only the prediction $\hat{y}_{0:\tau-1}$ as $\mathcal{P}^1=\left\{\hat{y}_{0:\tau-1}\right\}$. Furthermore, the distance measure in (\ref{eq:example_proto_setsequencepredictor})  is set as 
\begin{align}
    d({y}_{0:\tau-1},\hat{y}_{0:\tau-1})=\max_{t=0,\dots,\tau-1}w_t|y_{t}-\hat{y}_{t}|,
    \label{NC_weigthed}
\end{align}
 where $\{w_t\}^{\tau-1}_{t=0}$ are pre-determined positive coefficients. The weights can be chosen to be equal  \cite{stankeviciute2021conformal,lindemann2023conformal,dietterich2022conformal,lindemann2022safe}, or they may be decreasing over time to compensate for the fact that the prediction error is typically increasing in $t$ \cite{cleaveland2023conformal}.

The threshold $\lambda$ in (\ref{eq:example_proto_setsequencepredictor}) is selected based on a calibration data set $\mathcal{D}_{cal}=\{y_{-T:\tau-1}^i\}^n_{i=1}$ of trajectories drawn from the true distribution (\ref{joint}). To this end, TS-CP evaluates the distance between the deterministic prediction $\hat{y}^i_{0:\tau-1}$ and the corresponding $i$-th calibration data point $y^i_{0:\tau-1}$ as
\begin{align}
    d_i= d({y}^i_{0:\tau-1},\hat{y}^i_{0:\tau-1})
\end{align}
for all $i=1,\dots,n$. Then, following the general CP methodology \cite{vovk2005algorithmic}, it evaluates the  threshold $\lambda$ in (\ref{eq:example_proto_setsequencepredictor}) to equal the $\ceil{(n+1)(1-\alpha)}$-th smallest value of the calibration errors $\{d_i\}^n_{i=1}$ , i.e.,
\begin{align}
    \lambda^{TS-CP}=\mathcal{Q}_{1-\alpha}(\mathcal{D}_{cal})=\max\left\{d:\sum^n_{i=1}\mathds{1}\{d_i\leq d\}\leq \ceil{(n+1)(1-\alpha)}\right\}.
    \label{eq:quantile}
\end{align}
Accordingly, the TS-CP set predictor is given as
\begin{align}
\Gamma^{TS-CP}\left(y_{-T:-1}\right)=\left\{y_{0:\tau-1}\in \mathcal{Y}^{\tau}:d({y}_{0:\tau-1},\hat{y}_{0:\tau-1})\leq \mathcal{Q}_{1-\alpha}(\mathcal{D}_{cal})\right\}.
\label{standard_prediction_set}
\end{align}
Note that, by the choice of the distance in (\ref{NC_weigthed}), the per-time step predictor (\ref{eq:setsequencepredictor_at_time_t}), given by
\begin{align}
    \Gamma^{TS-CP}_{t}\left(y_{-T:-1}\right)=\left\{y\in \mathcal{Y}: |\hat{y}_{t}-y_{t}|\leq \frac{\mathcal{Q}_{1-\alpha}(\mathcal{D}_{cal})}{w_t}\right\},
    \label{standard_prediction_set_t}
\end{align}
 is centered on the predicted sample $\hat{y}_{t}$ with an error interval proportional to the empirical quantile $\mathcal{Q}_{1-\alpha}(\mathcal{D}_{cal})$. Furthermore, the set predictor $\Gamma^{TS-CP}\left(y_{-T:-1}\right)$ can be expressed as the Cartesian product of the per-time step intervals (\ref{standard_prediction_set_t}).

Assume that the calibration sequences in $\mathcal{D}_{cal}$ and the test sequence $y_{-T:\tau-1}$ are drawn i.i.d. from the distribution (\ref{joint}). Then, by the general properties of CP, the calibrated predictor (\ref{standard_prediction_set}) is guaranteed to include the true realization of the time series $y_{0:\tau-1}$ with a probability that is no smaller than $1-\alpha$ \cite{stankeviciute2021conformal,lindemann2023conformal,dietterich2022conformal,lindemann2022safe,cleaveland2023conformal}.  That is, TS-CP satisfies the coverage guarantee (\ref{eq:coverage_guarantee}), where the probability is evaluated with respect to the calibration and test sequences.

\section{Probabilistic Time Series Conformal Prediction }
\label{sec:ptscp}
The TS-CP set predictor $\Gamma_t^{TS-CP}\left(y_{-T:-1}\right)$ in (\ref{standard_prediction_set_t}) can only produce sets in the form of \emph{single real-valued} intervals. As such, TS-CP can become highly inefficient when the true distribution of the future evolution of the system, $p(y_{0:\tau-1}|y_{-T:-1})$, is multimodal (see Fig. \ref{fig:AoA}), and it  does not apply to discrete-valued time series. To address these limitations, in this section, we introduce PTS-CRC. Following Section \ref{sec:setting}, we differentiate between set predictors based on implicit and explicit probabilistic forecasters.   As we will see, explicit probabilistic predictors enable the definition of more general prediction schemes that may provide more informative set predictors.

\subsection{PTS-CRC via Implicit Sequence Models}

\begin{algorithm}
\caption{Probabilistic Time Series Conformal Risk Control (PTS-CRC)}\label{alg:ptscp}
\begin{algorithmic}
\Require Time series predictor $\hat{p}(y_{0:\tau-1}|y_{-T:-1})$, test input $y_{-T:-1}$, calibration data set $\mathcal{D}_{cal}=\{y^i_{-T:\tau-1}\}^n_{i=1}$,  maximum unreliability level $\alpha>0$, loss function $\mathcal{L}$, and integer $m>0$
\Ensure $\alpha$-reliable set predictor $\Gamma^{PTS-CRC}\left(y_{-T:-1}\right)$
\Statex \textit{//Offline Calibration Phase}
\ForAll{$y_{-T:\tau-1}^i\in \mathcal{D}_{cal}$}
\State sample predictions $\mathcal{P}^m_{i}=\{\hat{y}_{0:\tau-1}^j\}^m_{j=1}$ i.i.d. from the sequence model $\hat{p}({y}_{0:\tau-1}|y^i_{-T:-1})$
\EndFor
\State compute threshold $
    \lambda^{PTS-CRC}\leftarrow\inf\left\{\lambda: \sum^n_{i=1}\mathcal{L}\left(\Gamma_\lambda(\mathcal{P}^m_{i},y^i_{0:\tau-1})\right)+B\leq \alpha(n+1) \right\}.
    $
 \Statex \textit{// Testing Phase}
 \State sample predictions $\mathcal{P}^m=\{\hat{y}_{0:\tau-1}^j\}^m_{j=1}$ from the sequence model $\hat{p}({y}_{0:\tau-1}|y_{-T:-1})$
 \State obtain $\Gamma^{PTS-CRC}\left(y_{-T:-1}\right)=\bigcup_{\hat{y}_{0:\tau-1}\in\mathcal{P}^m}\left\{y_{0:\tau-1}\in \mathcal{Y}^{\tau}: d (\hat{y}_{0:\tau-1},{y}_{0:\tau-1})\leq \lambda^{PTS-CRC}\right\}$
\end{algorithmic}
\end{algorithm}
\label{sec:implicit}

Given past samples $y_{-T:-1}$, an implicit probabilistic predictor produces samples of predicted trajectories $\hat{y}_{0:\tau-1}$ from the predictive distribution $\hat{p}(y_{0:\tau-1}|y_{-T:-1})$, while not explicitly providing the value of the distribution $\hat{p}(\hat{y}_{0:\tau-1}|y_{-T:-1})$ for the generated sequences $\hat{y}_{0:\tau-1}$. Unlike TS-CP, which relies on a single predicted sequence $\hat{y}_{0:\tau-1}$, TS-CP leverages the capacity of a probabilistic sequence model to generate $m\geq1$ trajectories $\mathcal{P}^m=\{\hat{y}_{0:\tau-1}^j\}^m_{j=1}$ sampled i.i.d. from the model $\hat{p}({y}_{0:\tau-1}|y_{-T:-1})$. 

Based on the predicted trajectories $\mathcal{P}^m$, PTS-CRC applies the prototype-based set prediction (\ref{eq:example_proto_setsequencepredictor}) for a suitably designed threshold $\lambda^{PTS-CRC}$. Specifically, given a loss function $\mathcal{L}\left(\Gamma,y_{0:\tau-1}\right)$ and calibration data set $\mathcal{D}_{cal}=\{y^i_{-T:\tau-1}\}^n_{i=1}$ generated i.i.d. from the unknown distribution (\ref{joint}), the threshold $\lambda^{PTS-CRC}$ is chosen so as to guarantee the reliability constraint (\ref{eq:reliabilityguarantee}) for the target maximum unreliability level $\alpha$ under any loss function satisfying Assumption \ref{ass:loss}.

The reliability requirement (\ref{eq:reliabilityguarantee}) depends on the unknown joint distribution (\ref{joint}), and it can be estimated using the calibration data $\mathcal{D}_{cal}$. To this end, we evaluate the loss $\mathcal{L}\left(\Gamma_\lambda\left(y^i_{-T:-1}\right),y_{0:\tau-1}^i\right)$ for each $i$-th calibration data point by computing the set predictor $\Gamma_\lambda\left(y^i_{-T:-1}\right)$ in (\ref{eq:example_proto_setsequencepredictor}) based on  prototype predictions $\mathcal{P}_i^m=\{\hat{y}_{0:\tau-1,i}^j\}^m_{j=1}$ drawn from the sequence model. Note that we have made explicit the dependence of the set predictor (\ref{eq:example_proto_setsequencepredictor}) on the threshold $\lambda$.  Then, we evaluate the empirical average $1/n\sum^n_{i=1}\mathcal{L}\left(\Gamma_\lambda\left(y^i_{-T:-1}\right),y_{0:\tau-1}^i\right)$ by averaging over the calibration data set. Intuitively, PST-CRC chooses the threshold $\lambda^{PTS-CRC}$ in such a way that this empirical estimate is no larger than $\alpha$. More precisely, we have
\begin{align}
    \lambda^{PTS-CRC}:=\inf\left\{\lambda: \frac{1}{n+1}\left(\sum^n_{i=1}\mathcal{L}\left(\Gamma_\lambda\left(y^i_{-T:-1}\right),y_{0:\tau-1}^i\right)+B\right)\leq \alpha \right\},
    \label{right_lambda}
\end{align}
where the empirical estimate of constraint (\ref{eq:reliabilityguarantee}) is corrected by adding  a fictitious $(n+1)$-th data point with maximal loss value $B$ (see Assumption \ref{ass:loss}). As explained in Section VI, this correction follows the CRC framework.

The PTS-CRC procedure, producing PTS-CRC predicted set 
\begin{align}
    \Gamma^{PTS-CRC}\left(y_{-T:-1}\right)=\hspace{-1em}\bigcup_{\hat{y}_{0:\tau-1}\in\mathcal{P}^m}\left\{y_{0:\tau-1}\in \mathcal{Y}^{\tau}: d (\hat{y}_{0:\tau-1},{y}_{0:\tau-1})\leq \lambda^{PTS-CRC}\right\},
\label{implicit_prediction_set_norm_t}
\end{align}
is summarized in Algorithm \ref{alg:ptscp}. PTS-CRC satisfies the following reliability guarantee.

\begin{theorem}
\label{theorem1}
Assuming that the samples in the calibration data set $\mathcal{D}_{cal}$ and the test sample $y_{-T:\tau-1}$ are i.i.d. from distribution (\ref{joint}), and that the loss function $\mathcal{L}\left(\Gamma,y_{0:\tau-1}\right)$ satisfies Assumption \ref{ass:loss}, the PTS-CRC set predictor $\Gamma^{PTS-CRC}\left(y_{-T:-1}\right)$ in (\ref{implicit_prediction_set_norm_t})  
satisfies the $\alpha$-reliability guarantee (\ref{eq:reliabilityguarantee}), where the expectation is taken with respect to the calibration data set $\mathcal{D}_{cal}$, the test data point $y_{-T:\tau-1}$ and the prototypes $\left\{\mathcal{P}^m,\left\{\mathcal{P}^m_i\right\}^n_{i=1}\right\}$, with the latter drawn i.i.d. form the respective predictive distributions $\left\{\hat{p}(y_{0:\tau-1}|y_{-T:-1}),\left\{\hat{p}(y_{0:\tau-1}|y^i_{-T:-1})\right\}^n_{i=1}\right\}$.
\label{th:reliability_implicit}
\end{theorem}

The properties of PTS-CRC stated in Theorem \ref{theorem1} can be proved by leveraging tools from the theory of CRC \cite{angelopoulos2022conformal} and probabilistic CP \cite{wang2022probabilistic}. This is  discussed in Section \ref{sec:derivation}.

\subsection{PTS-CRC via Explicit Sequence Models}
\label{sec:explicit}

In this subsection, we propose E-PTS-CRC, a variant of PTS-CRC that leverages \emph{explicit} probabilistic predictors. As detailed in Section \ref{sec:setting}, explicit forecasters not only allow a trajectory $\hat{y}_{0:\tau-1}$ to be sampled from the model distribution $\hat{p}(y_{0:\tau-1}|y_{-T:-1})$, but they also provide the value of the distribution $\hat{p}(\hat{y}_{0:\tau-1}|y_{-T:-1})$ for the synthesized sample.
We take inspiration from the literature on language models \cite{fan2018hierarchical}, with the aim of generating sets of predicted trajectories that are better representatives of the plausible evolutions of the input sequence $y_{0:\tau-1}$.

This objective is accomplished via a biased sampling procedure that generates samples $\hat{y}_{0:\tau-1}$ from a distribution $\hat{q}({y}_{0:\tau-1}|y_{-T:-1})$ that is generally distinct from the sequence model $\hat{p}(y_{0:\tau-1}|y_{-T:-1})$, satisfying additional desirable properties.  For instance, in the context of text generation, which corresponds to a time series forecasting problem over a sequence of words, trajectories with the largest likelihood are often nonsensical \cite{holtzman2019curious}, and hence one may wish to filter out sequences by typicality rather than likelihood \cite{meister2023locally}. Furthermore, it may be desirable to explicitly avoid the generation of sequences that are too unlikely. As exemplary strategies, we elaborate here on sequence-level filtering \cite{wang2022probabilistic} and autoregressive filtering \cite{fan2018hierarchical,holtzman2019curious,meister2023locally}.

\subsubsection{Sequence-level filtering}
Sequence-level filtering aims at obtaining samples from high-density regions of the predictive distribution, while reducing the occurrence of unlikely trajectories. This is done by filtering out samples with low likelihood from a set of trajectories sampled from the predictive distribution $\hat{p}(y_{0:\tau-1}|y_{-T:-1})$ \cite{wang2022probabilistic}. Specifically, given an explicit model $\hat{p}({y}_{0:\tau-1}|y_{-T:-1})$, one samples a set ${\mathcal{P}}^{\ceil{m(1+\kappa)}}$  of $\ceil{m(1+\kappa)}$ trajectories obtained i.i.d. from $\hat{p}(y_{0:\tau-1}|y_{-T:-1})$ given some $\kappa > 0$, and then obtains a subset ${\mathcal{P}}^{m}\subset {\mathcal{P}}^{\ceil{m(1+\kappa)}}$ by selecting the $m$ trajectories with the largest distribution value from set ${\mathcal{P}}^{\ceil{m(1+\kappa)}}$.

\subsubsection{Autoregressive  filtering}
In autoregressive filtering, sample selection is done on a per-time step basis. For example, in \emph{top-$k$ sampling} \cite{fan2018hierarchical}, which applies to discrete sets $\mathcal{Y}$, the next sample $\hat{y}_t$ is constrained to lie within the set $\mathcal{S}_k$ of samples $y_t$ with the top-$k$ largest distribution value $\hat{p}(y_{t}|\hat{y}_{0:t-1},y_{-T:-1})$. Sampling is  hence done from a truncated probability $\hat{q}(y_{t}|\hat{y}_{0:t-1},y_{-T:-1})\propto\hat{p}(y_{t}|\hat{y}_{0:t-1},y_{-T:-1})\mathds{1}(y\in\mathcal{S}_k)$. Other examples include $p$-nucleus sampling \cite{holtzman2019curious} and locally typically sampling \cite{meister2023locally}, which respectively apply to continuous and discrete sets $\mathcal{Y}$.

\section{ PTS-CRC Model Predictive Control}
\label{sec:mpc}

In this section, we introduce open-loop and closed-loop policies for the MPC problem (\ref{mpc}) by leveraging PTS-CRC to predict the target process trajectories. Control policies based on TS-CP, which was reviewed in Section \ref{background:tscp}, were presented in \cite{lindemann2022safe}, and they will be obtained as a special case of the more general framework put forth here.

\subsection{Open-Loop MPC}
We first consider the open-loop MPC control problem (\ref{mpc}), whereby the action $u_t$ is allowed to depend only on the initial state $s_0$, the past samples $y_{-T:-1}$, and the (implicit or explicit) probabilistic predictor $\hat{p}(y_{0:\tau-1}|y_{-T:-1})$. As discussed in Section \ref{sec:mpc_intro}, meeting the average cost constraint (\ref{mpc_constraint}) is made complicated by the fact that the distribution $p(y_{0:\tau-1}|y_{-T:-1})$ of the target process $y_{0:\tau-1}$ is unknown. In order to gauge the impact of the prediction errors on the performance of a control policy in terms of the constraint (\ref{mpc_constraint}), we introduce the following assumption, which limits the sensitivity of the constraint to changes in the target process.

\begin{assumption}[Constraint Sensitivity]
    \label{ass:lipshitz}
    For some $L>0$, the constraint function $c(s_{0:\tau-1},y_{0:\tau-1})$ is $L$-Lipschitz in the second argument $y_{0:\tau-1}\in\mathcal{Y}^\tau$ with respect to some metric $m:\mathcal{Y}^\tau\times \mathcal{Y}^\tau \to \mathbb{R}$ in the space of trajectories. That is, we have the inequality
    \begin{align}
        |c(s_{0:\tau-1},y'_{0:\tau-1})-c(s_{0:\tau-1},y''_{0:\tau-1})|\leq L m(y'_{0:\tau-1},y''_{0:\tau-1})
        \label{eq:lipshitz}
    \end{align}
    for all sequences $s_{0:\tau-1}$ and all pairs of target process trajectories $y'_{0:\tau-1}$ and $y''_{0:\tau-1}$.
\end{assumption}

Assumption \ref{ass:lipshitz} states that we can identify a function $m(\cdot,\cdot)$ with the property that switching between any two target sequences $y'_{0:\tau-1}$ and $y''_{0:\tau-1}$ cannot change the constraint $c(\cdot,\cdot)$ by more than $L m(y'_{0:\tau-1},y''_{0:\tau-1})$ for some constant $L>0$.

For example, the constraint function 
\begin{align}
    c(s_{0:\tau-1},y'_{0:\tau-1})=\left(\sum^{\tau-1}_{t=0} |s_{t}-y'_{t}|^p\right)^{1/p}
\end{align}
for $p\geq 1$, which evaluates the $\ell_p$ distance between the state sequence $s_{0:\tau-1}$ and the target process $y'_{0:\tau-1}$, satisfies Assumption \ref{ass:lipshitz} with $L=1$ for the function $m(\cdot,\cdot)=c(\cdot,\cdot)$. Another constraint function is 
\begin{align}
    c(s_{0:\tau-1},y'_{0:\tau-1})=\sum^{\tau-1}_{t=0} \log(1+s_{t}y'_{t}),
\end{align}
which will be seen in Section \ref{sec:experiments} to be relevant for the control of wireless systems. This constraint function can be shown to satisfy Assumption \ref{ass:lipshitz} with $L=\max_{s\in\mathcal{S}}|s|$ and function $ m(y'_{0:\tau-1},y''_{0:\tau-1})=\sum^{\tau-1}_{t=0} |y'_{t}-y''_{t}|$, assuming that the inequality $s_t y_t \geq 0$ holds for any state $s_t$ and target sample $y_t$.

Under Assumption \ref{ass:lipshitz}, we now derive a \emph{surrogate MPC problem} whose feasibility set is guaranteed to be a subset of the feasibility set of the original control problem (\ref{mpc}).
 In other words, a solution to the surrogate problem is also guaranteed to be feasible for the original problem, and hence the surrogate problem imposes more conservative constraints. As we will discuss, the key advantage of the surrogate MPC problem is that it can be addressed by using the available predictor, while the original problem (\ref{mpc}) is not accessible given the dependence of the constraint (\ref{mpc_constraint}) on the unknown target process distribution.
 \begin{theorem}[PTS-CRC-based surrogate open-loop MPC problem]
\label{th:mpc_exp_open_loop}
Consider the PTS-CRC predictor $\Gamma^{PTS-CRC}\left(y_{-T:-1}\right)$ in (\ref{right_lambda})-(\ref{implicit_prediction_set_norm_t})  obtained with distance measure $d(\cdot,\cdot)=m(\cdot,\cdot)$ for some target reliability level $\alpha>0$ and with the loss
\begin{align}
    \mathcal{L}(\Gamma,y_{0:\tau-1})=\min_{\tilde{y}_{0:\tau-1}\in\Gamma} m(y_{0:\tau-1},\tilde{y}_{0:\tau-1}).
    \label{eq:control_loss}
\end{align}
Any solution to the problem
\begin{subequations}
\begin{align}
\minimize_{u_{0}\dots,u_{\tau-1}} \quad & \sum^{\tau-1}_{t=0} J(s_t,u_t)\\
\textup{s.t.} \quad & s_t=f(s_{t-1},u_t), \ \textup{ for }\ t=0,\dots,\tau-1\\
&c(s_{0:\tau-1},\hat{y}_{0:\tau-1})\leq  -L\alpha   \ \textup{ for all }\ \hat{y}_{0:\tau-1}\in \Gamma^{PTS-CRC}\left(y_{-T:-1}\right)
\label{eq:surrogate_mpc_constraint}
\end{align}
\label{eq:surrogate_mpc}%
\end{subequations}
yields feasible solutions also for the original MPC problem (\ref{mpc}) in which the average reliability constraint (\ref{mpc_constraint}) is evaluated on average with respect to the evolution $y_{0:\tau-1}$, the calibration sequences $\mathcal{D}_{cal}$, and the prototypes $\left\{\mathcal{P}^m,\left\{\mathcal{P}^m_i\right\}^n_{i=1}\right\}$, with the latter drawn i.i.d. form the respective predictive distributions $\left\{\hat{p}(y_{0:\tau-1}|y_{-T:-1}),\left\{\hat{p}(y_{0:\tau-1}|y^i_{-T:-1})\right\}^n_{i=1}\right\}$ .
\end{theorem}

Theorem \ref{th:mpc_exp_open_loop} justifies the adoption of the surrogate problem (\ref{eq:surrogate_mpc}) in lieu of the problem (\ref{mpc}). In problem (\ref{eq:surrogate_mpc}),  constraint (\ref{mpc_constraint}) is relaxed by taking the expectation not only with respect to target process $y_{0:\tau-1}$, but also over the prediction $\Gamma^{PTS-CRC}\left(y_{-T:-1}\right)$. The relaxed constraint (\ref{eq:surrogate_mpc_constraint}) becomes more stringent as the Lipschitz constant $L$ increases, indicating, by (\ref{eq:lipshitz}), that the constraint becomes more sensitive to changes in the target process.
Since (\ref{eq:surrogate_mpc_constraint}) is a more stringent requirement as compared to (\ref{mpc_constraint}), there exists scenarios in which the original MPC problem (\ref{mpc}) is feasible but the surrogate problem  
(\ref{eq:surrogate_mpc}) is not. The reliability threshold $\alpha$ in constraint (\ref{eq:surrogate_mpc_constraint}), which may be freely chosen, dictates the trade-off between the size of the search space $\Gamma^{PTS-CRC}\left(y_{-T:-1}\right)$ and the strictness of the inequality. 


The MPC controller based on TS-CP proposed in \cite{lindemann2022safe} can be recovered as a special case of the PTS-CRC-based control policy of Theorem \ref{th:mpc_exp_open_loop}. In particular, given a reliability level $\delta>0$, the controller in \cite{lindemann2022safe} addresses constraints (\ref{mpc_constraint}) of the form 
\begin{align}
    \mathbb{E}\left[\mathds{1}\left\{c(s_{0:\tau-1},y_{0:\tau-1})>0\right\}\right]\leq \delta,
\end{align}
by leveraging the TS-CP predictor (\ref{standard_prediction_set}).

\subsection{Closed-Loop MPC}
In the closed-loop setting, at every time step $t$,  as detailed in Section \ref{sec:mpc_intro}, the controller receives a feedback signal providing the current value of the state of the target process $y_t$. As such, the control $u_t$ at time $t$ is allowed to depend on the observed sequence $y_{-T:t-1}$, the sequence of state $s_{0:t-1}$, and the probabilistic predictor $\hat{p}(y_{t:\tau-1}|y_{-T:t-1})$. 

In this setting, the control sequence $u_0,...,u_{\tau-1}$ is designed by following a \emph{receding horizon} strategy. Accordingly, at every time step $t>0$, the control action $u_t$ is obtained by optimizing the future control sequence $u_t,...,u_{\tau-1}$ and then retaining only the first action. As we describe next, this optimization leverages leverages the output of the PTS-CRC predictor $\Gamma^{PTS-CRC}(y_{-T:t-1})$ and the feedback sequence $y_{0:t-1}$ in a manner similar to Theorem \ref{theorem1}. 


Under Assumption \ref{ass:lipshitz}, at each time step $t=0,\dots,\tau-1$, based on the observed sequence $y_{-T:t-1}$ we define a \emph{surrogate MPC problem} whose feasibility set is guaranteed to be a subset of the feasibility set of the control problem (\ref{mpc}) for the time interval $t,\dots,\tau-1$.
The surrogate problem imposes more conservative constraints, which be addressed using the available predictor $\Gamma^{PTS-CRC}(y_{-T:t-1})$.

 \begin{theorem}[PTS-CRC-based closed-loop MPC]
\label{th:mpc_exp_closed_loop}
For each time step $t=0,\dots,\tau-1$, consider the PTS-CRC predictor $\Gamma^{PTS-CRC}(y_{-T:t-1})$ obtained with the distance measure $d(\cdot,\cdot)=m(\cdot,\cdot)$ for some target reliability level $\alpha>0$ and calibrated via (\ref{right_lambda}) on the loss
\begin{align}
    \mathcal{L}(\Gamma,y_{0:\tau-1})=\min_{\tilde{y}_{0:\tau-1}\in\Gamma} m(y_{0:\tau-1},\tilde{y}_{0:\tau-1}).
    \label{eq:control_loss_2}
\end{align}
Then, the sequence of actions $u_0,\dots,u_{\tau-1}$, in which $u_t$ is obtained as a solution of 
\begin{subequations}
\begin{align}
\minimize_{u_{t}\dots,u_{\tau-1}} \quad & \sum^{\tau-1}_{k=t} J(s_k,u_k)\\
\textrm{s.t.} \quad & s_k=f(s_{k-1},u_k), \ \textup{ for }\ k=0,\dots,\tau-1\\
&c(s_{0:\tau-1},\hat{y}_{0:\tau-1})\leq  -L\alpha ,  \ \textup{ for all }\ \hat{y}_{0:\tau-1}\in \Gamma^{PTS-CRC}\left(y_{-T:t-1}\right)  , 
\label{eq:surrogate_mpc_closedloop_constraint}
\end{align}
\label{eq:surrogate_mpc_closedloop}%
\end{subequations}
yields feasible solutions also for the original MPC problem (\ref{mpc}) in which the average reliability constraint is evaluated on average with respect to prediction $\{\Gamma^{PTS-CRC}\left(y_{-T:t-1}\right)\}^{\tau-1}_{t=0}$ and the evolution $y_{0:\tau-1}$. 
\end{theorem}

In a manner similar to the open-loop control case studied in the previous subsection, the closed-loop control policy based on TS-CP \cite{lindemann2022safe} can be recovered as a special instantiation of the PTS-CRC policy of Theorem \ref{th:mpc_exp_closed_loop}.

\section{Connecting PTS-CRC with CP, Probabilistic CP, and CRC}
\label{sec:derivation}

In this section, we first briefly review CP, probabilistic CP (PCP), and CRC, and then we describe PTS-CRC as a novel application of the principles underlying PCP and CRC to time series data.

\subsection{Conformal Prediction}

CP, PCP, and CRC apply to a general supervised learning setting in which data points take the form of pairs $(x,y)$ with input $x$ and output $y\in \mathcal{Y}$. These schemes assume the availability of a  calibration data set $\mathcal{D}_{cal}=\{(x,y)_i\}^n_{i=1}\in\left(\mathcal{X}\times\mathcal{Y}\right)^n $ of input-output pairs $(x,y)$, which are assumed to be jointly distributed with the test pair in an   i.i.d. manner. The theory also generalizes directly to exchangeable data points \cite{vovk2005algorithmic}. 

CP transforms a pre-designed predictor $\hat{y}=f(x)$ into a set predictor $\Gamma^{CP}(x)\subseteq\mathcal{Y}$ that contains the true output $y$ with any target probability $1-\alpha$, with probability evaluated with respect to calibration and test data.  
The CP set predictors $\Gamma^{CP}(x)$ depends on the choice of a \emph{non-conformity (NC) scoring function} $d(\cdot):\mathcal{Y}\times \mathcal{Y}\to \mathbb{R}$ that measures the extent to which model's prediction conforms with the ground truth $y$. Specifically, for each data point $(x,y)$, the NC score is evaluated as  $d(f(x),y)$, and we denote as $d_i=d(f(x_i),y_i)$ the NC score for the $i$-th calibration data point.  Then, the CP set predictor is defined as
\begin{align}
    \Gamma^{CP}(x)=\{y:d(f(x),y)\leq Q_{1-\alpha}(\mathcal{D}_{cal})\},
    \label{standard_CP_set_predictor}
\end{align}
where $Q_{1-\alpha}(\mathcal{D}_{cal})$ is the $\ceil{(n+1)(1-\alpha)}$-th smallest value of the set $\{d_i\}^n_{i=1}$ of calibration NC scores. It can be shown that (\ref{standard_CP_set_predictor}) satisfies the \emph{coverage} guarantee \cite{vovk2005algorithmic}.
\begin{align}
    \Pr\left[y\notin\Gamma^{CP}(x)\right]\leq \alpha.
    \label{eq:cpcoverge}
\end{align}

The TS-CP set predictor (\ref{standard_prediction_set}) can be obtained as an application of the CP set predictor by considering $x$ to be the past samples $y_{-T:-1}$ and the target $y$ the future samples $y_{0:\tau-1}$, while using as the NC scoring function the maximum per-sample weighted error (\ref{NC_weigthed}).

\subsection{Probabilistic Conformal Prediction}
PCP is a variant of CP that aims at producing discontinuous prediction sets based on samples obtained from probabilistic predictors. Specifically, given a probabilistic predictor $\hat{p}(y|x)$ and an input $x$, PCP generates $m$  i.i.d. predictions $\mathcal{P}^m(x)=\{\hat{y}_i\}^m_{i=1}$. Then, given an  NC scoring function $d(\cdot):\mathcal{Y}\times \mathcal{Y}\to \mathbb{R}$,  it produces the predictive set
\begin{align}
\Gamma^{PCP}(x)=\bigcup_{\hat{y}\in\mathcal{P}^m(x)}\left\{y:d(y,\hat{y})\leq \lambda\right\},
\end{align}
where the threshold $\lambda>0$ is  selected as the  $\ceil{(n+1)(1-\alpha)}$-th smallest value of the set $\{d_i\}^n_{i=1}$ of calibration NC scores $d_i=\min_{\hat{y}\in\mathcal{P}^m(x)}d(y_i,\hat{y})$. PCP can be shown to also satisfy the coverage condition (\ref{eq:cpcoverge}).

In the special case of a the miscoverage loss (\ref{eq:miscoverageloss}) and assuming implicit probabilistic predictors, in a manner similar to TS-CP, PTS-CRC can be thought of as an application of PCP to time series prediction.

\subsection{Conformal Risk Control}
\label{sec:CRC}
CRC is a generalization of CP that addresses more general reliability requirements, beyond the the coverage guarantee (\ref{eq:cpcoverge}) \cite{angelopoulos2022conformal}. Given a \emph{deterministic} predictor $\hat{y}=f(x)$, calibration data $\mathcal{D}_{cal}$, and test input $x$, CRC produces a set predictor $\Gamma^{CRC}(x)$ that satisfies the average constraint \begin{equation}\mathbb{E}[\mathcal{L}(\Gamma^{CRC}(x),y)]\leq \alpha\end{equation} for a bounded loss $\mathcal{L}\left(\Gamma,y\right)$.  To this end, the CRC set predictors is given by
\begin{align}
    \Gamma^{CRC}(x)=\{y:d(f(x),y)\leq \lambda^{CRC}\},
    \label{general_CRC_set_predictor}
\end{align}
where the threshold $\lambda^{CRC}$ is chosen so as to ensure the inequality $1/(n+1)(\sum^n_{i=1}\mathcal{L}(\Gamma_\lambda(x^i,y^i)+B)\leq \alpha$, where $B$ is a bound on the loss function and we have defined  $\Gamma_\lambda(x)=\{y:d(f(x),y)\leq \lambda\}$.

In the special case of deterministic predictors ($m=1$), PTS-CRC can be seen as an application of CRC to time series prediction. Overall, in order to capture forking uncertainties in time series prediction while accounting for general loss functions, PTS-CRC borrows from PCP the idea of relying on multiple stochastic predictions, and for CRC the idea of calibrating a set prediction on the basis of an empirical estimate of the loss function. Furthermore, in order to enhance the predictive efficiency, PTS-CRC integrates the use of explicit probabilistic  predictors, leveraging recent work on sequence modeling.

\section{Experiments}
\label{sec:experiments}
In this section, we explore the application of PTS-CRC set predictors in the context of wireless networking. To begin, we address the challenge of reliably forecasting the evolution of the channel gain between a base station and users moving in an urban cell scenario. Subsequently, we harness the predicted channel behavior to develop model predictive power control policies subject to interference and energy efficiency requirements.
\subsection{Simulation Scenario}

\begin{figure}[t]
     \centering
     \begin{subfigure}[b]{0.45\textwidth}
         \centering
         \includegraphics[width=\textwidth]{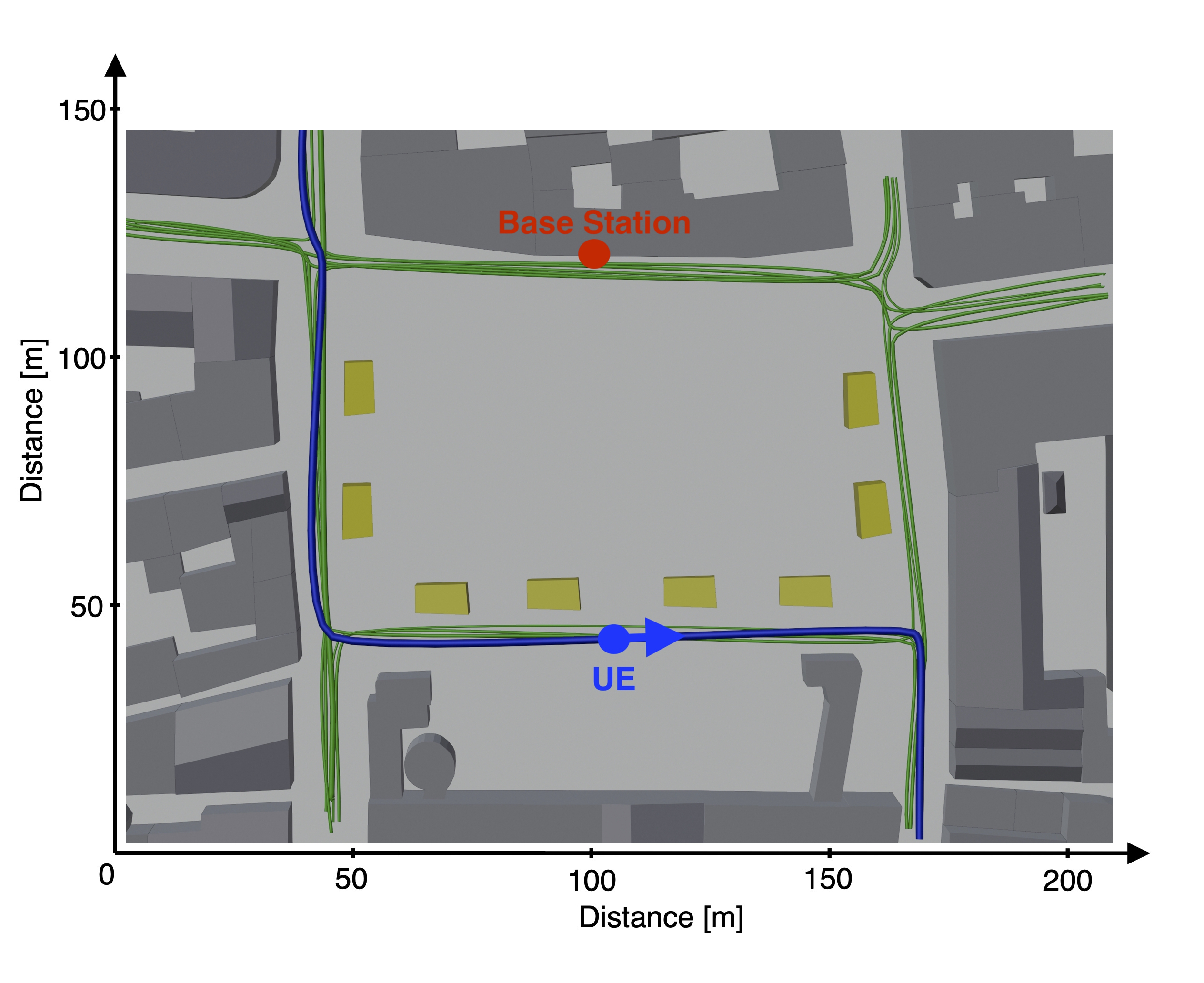}
         \caption{Deployment Area \label{fig:deployment}}
     \end{subfigure}
     \hspace{2em}
     \begin{subfigure}[b]{0.32\textwidth}
         \centering
         \includegraphics[width=1.1\textwidth]{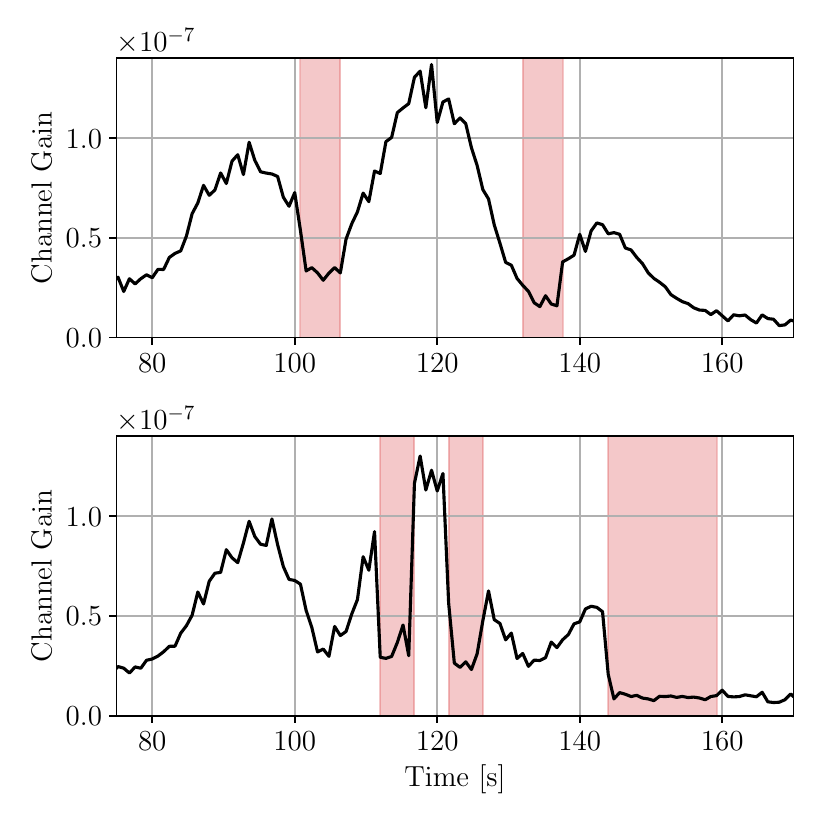}
         \caption{Channel gain evolutions\label{fig:trace}}
     \end{subfigure}
    \caption{ (a) Top-view of the simulation scenario: The base station (BS), represented by a red circle, serves a user equipment (UE), blue circle, that moves along one of the possible 30 paths in green. The scenario includes static obstacles such as buildings, represented as dark grey blocks, as well as dynamic obstacles that may or may not be present, in yellow. (b) Two possible channel gain sequence realizations for a UE moving along the blue path shown in Fig.\ref{fig:deployment}. The red-shaded areas correspond to random blockage events during which the line-of-sight (LoS) component is blocked due to the presence of dynamic obstacles.}
       \label{fig:scene} 
\end{figure}

We consider the urban microcell deployment depicted in Fig. \ref{fig:deployment}. In Marienhof square, located in Munich, a base station (BS) is located at the top of a building facing the square. The BS serves users that move across the square with a fixed constant speed of $1.5$ m/s following one of 30 possible trajectories. All trajectories are equally likely. As illustrated in Fig. \ref{fig:deployment}, obstacles in the scene can obstruct the line of sight (LoS) component between the BS and the user equipment (UE). Each obstacle, shown in Fig. \ref{fig:deployment}, can be present or not, independently from the other obstacles, with $P_{b}$, which we set as $P_b=0.5$ (see, e.g. \cite{haneda20165g}).

The BS and the UE communicate using a single receiving and transmitting antenna system operating at a center frequency $f_c=2.14$ GHz with a bandwidth $B=120$ KHz. The wireless channel is simulated using the ray-tracing simulator Sionna RT \cite{sionna}. Accordingly, the baseband channel impulse response is obtained by simulating the wave propagation of the transmitted signal, and it is described by the superposition of $N_R$ rays as 
\begin{align}
    h(\tau)=\sum^{N_R}_{i=1}a_ie^{-j2f_c\pi\tau_i}\delta(\tau-\tau_i),
\end{align}
where $a_i\in \mathbb{C}$ and $\tau_i\in \mathbb{R}$ are the channel complex coefficients and the delay associated with the $i$-th simulated ray. Assuming that the delay spread $\Delta_\tau=\max_{i,j}|\tau_i-\tau_i|$ is small compared to the symbol time $1/B$, the channel gain is evaluated as 
\begin{align}
    g=\left|\sum^{N_{rays}}_{i=1}a_ie^{-j2f_c\pi\tau_i} \right|^2.
    \label{eq:channel_gain}
\end{align}
We assume that the BS has a maximum transmit power of $P_{\textrm{max}}=1$ W and that the communication link is affected by additive white Gaussian noise with a noise spectral density $N_0=10^{-15}$ W/Hz.
 
For every user in the cell, the value of the channel gain (\ref{eq:channel_gain}) is estimated at the UE based on a reference signal that is periodically transmitted by the BS with a periodicity $T_s=80$ ms. The UE evaluates an average over 10 measurements of the channel gain, and the average is sent back to the BS \cite{dahlman20164g}. Accordingly, the BS receives a channel gain estimate $g_t$ every 800 ms. As exemplified in Fig. \ref{fig:trace}, the evolution of the time series $g_t$ depends on the path followed by the UE as well as on the random blockage events.

\subsection{Reliable Channel Gain Prediction}
\begin{figure}[t]
     \centering
     \begin{subfigure}[b]{0.4\textwidth}
         \centering
         \includegraphics[width=\textwidth]{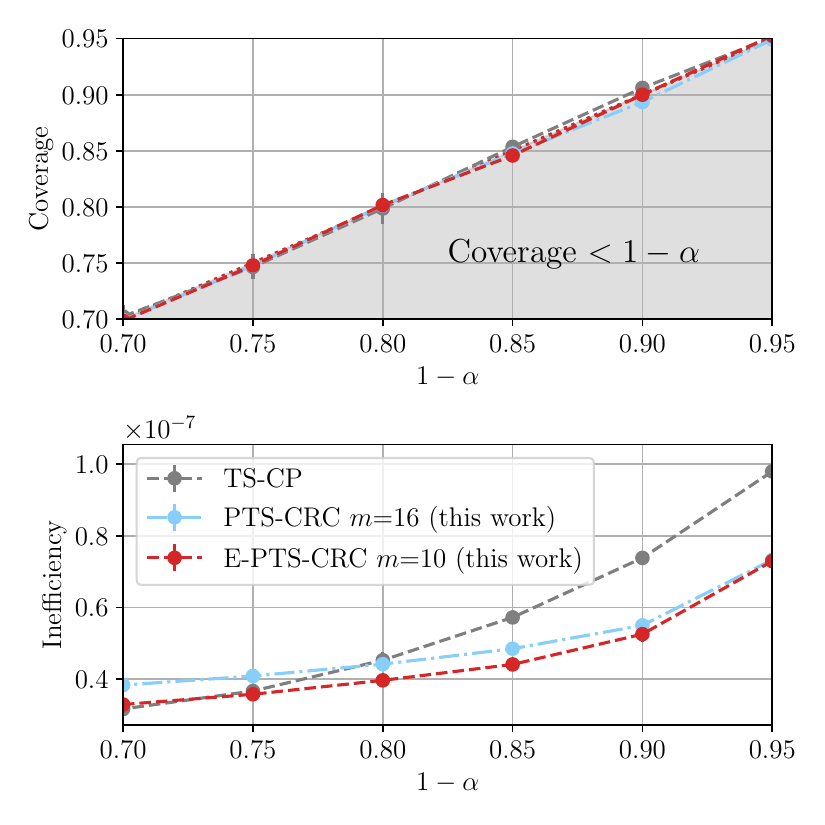}
         \caption{Coverage and Inefficiency\label{fig:cov_and_eff}}
     \end{subfigure}
     \hspace{2em}
     \begin{subfigure}[b]{0.4\textwidth}
         \centering
         \includegraphics[width=\textwidth]{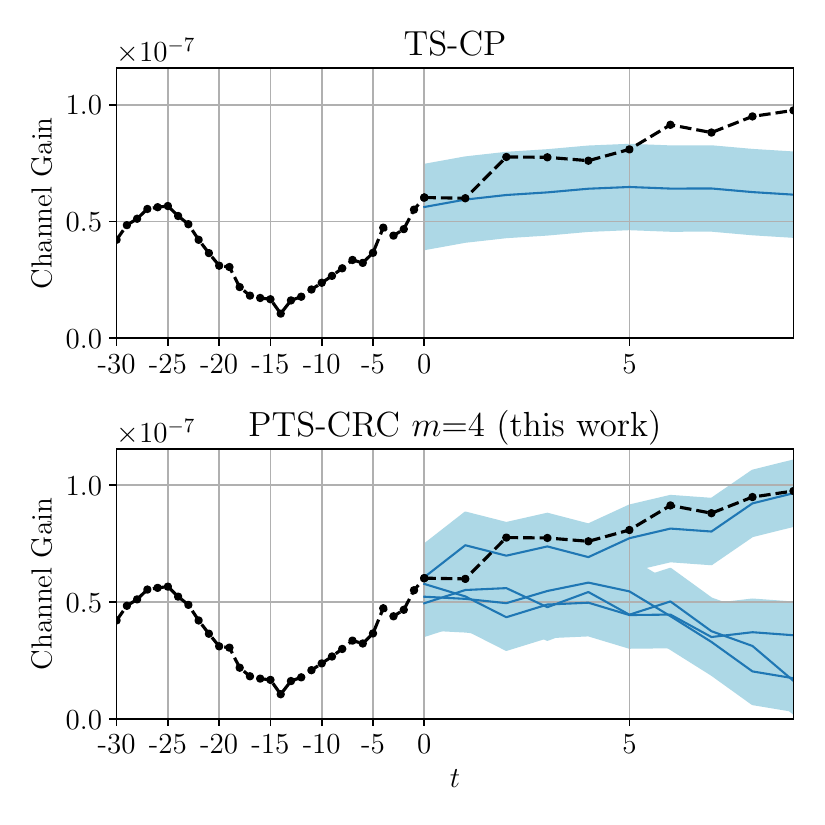}
         \caption{Prediction Examples\label{fig:pred_example}}
     \end{subfigure}
    \caption{(a) Test coverage probability \ref{eq:coverage_guarantee} and inefficiency \ref{inefficiency} of TS-CP \cite{stankeviciute2021conformal}, PTS-CRC with $m=16$ prediction samples and E-PTS-CRC. (b) Example of set predictions produced by TS-CP and PTS-CRC based on $m=4$ prototypes.  }
       \label{fig:coverage_channel_prediction} 
\end{figure}

As a first task, we address the problem of reliably forecasting the future evolution of the channel gain at the BS based on past feedback messages received from a UE in the cell. Specifically, given the past $T$ channel gain estimates $g_{-T:-1}$, our objective is to generate a set predictor that includes the true future evolution $g_{0:\tau-1}$ with a probability no smaller than $1-\alpha$.

To achieve this goal, we train a DeepAR probabilistic forecaster $\hat{p}(g_{0:\tau-1}|g_{-T:-1})$ \cite{salinas2020deepar} using a training dataset comprising 73k channel gain sequences recorded from UEs moving within the deployment area as explained in the previous subsection.

We explore different calibration strategies to transform the trained forecaster into a reliable set predictor. Our options include TS-CP \cite{stankeviciute2021conformal}, which involves applying CP to the average prediction obtained from the predictive distribution of the DeepAR model, i.e., $f(g_{-T:-1})=\mathbb{E}\left[\hat{p}(g_{0:\tau-1}|g_{-T:-1})\right]$. Additionally, we consider the proposed PTS-CRC (Sec. \ref{sec:implicit}), obtained by directly sampling $m=16$ prototypes from the probabilistic predictor $\hat{p}(g_{0:\tau-1}|g_{-T:-1})$, and the proposed E-PTS-CRC (Sec. \ref{sec:explicit}) based on sequence-level filtering, constructed on the subset of $m=10$ prototypes with highest likelihood from the original set of 16 prototypes. Calibration uses a data set of $N_{cal}=1000$ time series, and Fig. \ref{fig:cov_and_eff} presents the test coverage levels and test efficiency of the set predictors averaged over $N_{te}=1000$ independently generated test time series.

As depicted in the top panel of Fig. \ref{fig:cov_and_eff}, all calibration methods produce set predictors that meet the desired target coverage levels $1-\alpha$. The coverage probability, defined in (\ref{eq:coverage_guarantee}), evaluates the fraction of test trajectories that lie within the predicted set. As seen as in the bottom panel of Fig. \ref{fig:cov_and_eff}, as the coverage requirement $1-\alpha$ increases, CP yields sets with lower efficiency compared to the proposed the proposed PTS-CRC and E-PTS-CRC. The inefficiency is measured by the average size of the predicted set per time instant. As illustrated with two examples of predictions in Fig. \ref{fig:pred_example}, the higher efficiency of the proposed probabilistic set predictors can be attributed to their ability to output disjoint sets that better capture the multimodal residual uncertainty associated with unknown mobility patterns and blockage events.
\subsection{Open-Loop Model Predictive Power Control for Interference Mitigation}
\begin{figure}[t]
     \centering
     \includegraphics[width=0.5\textwidth]{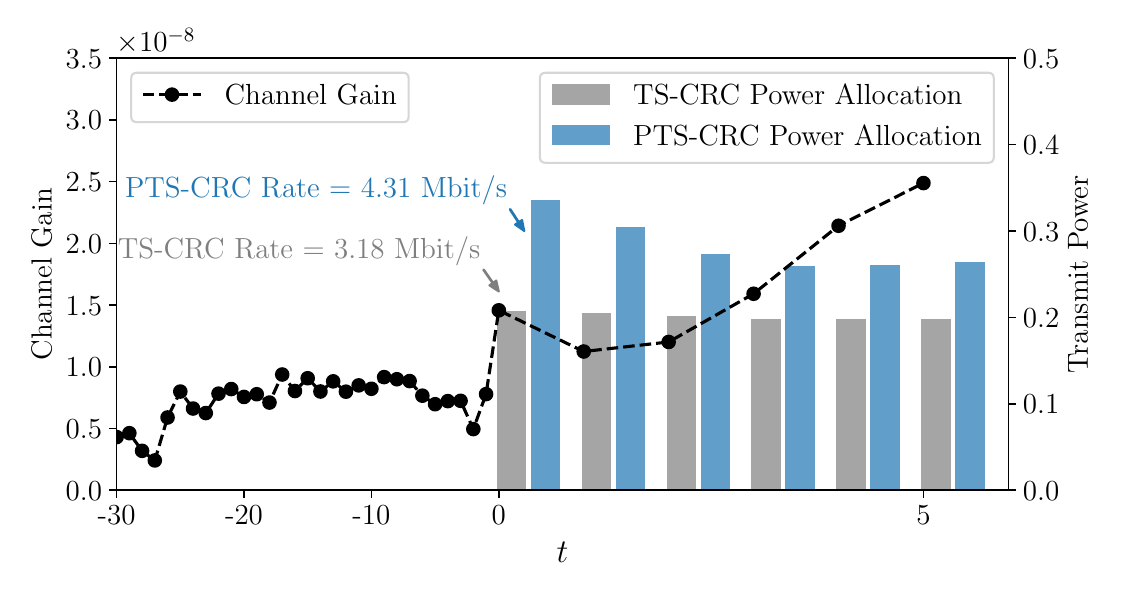}
     \caption{Power allocation example obtained solving the model predictive power control problem based on the TS-CRC \cite{angelopoulos2022conformal} and the proposed PTS-CRC predictor. The true channel realization (unknown) is shown as a dashed black line. Both the TS-CRC (in grey) and the PTS-CRC (in blue) power allocations ensure that the maximum cumulative interference over $k=3$ slots does not exceed the safety threshold $\gamma$. However, the larger efficiency of the PTS-CRC predictor translates into a rate that is 33\% larger than the one obtained using the TS-CRC predictor.}
     \label{fig:example_pc}
\end{figure}

In this subsection we leverage the reliable channel gain set predictors evaluated above to derive model predictive power control policies that satisfy interference constraints. More specifically, we consider the scenario in which licensed users (LU) and unlicensed users (UU) coexist within the cell, and the BS is tasked with the problem of modulating its transmit power $P_0,\dots,P_{\tau-1}$ over the next $\tau$ communication slots in order to maximize the sum-rate of the UU, while controlling the interference experienced by the LU. This formulation is motivated by the fact that UUs are typically served in a best-effort fashion whereas higher priority LUs have strict reliability requirements (see, e.g., \cite{hoven2005power,sahai2006fundamental}). 

The BS observes the past evolution of the channel gain  $g^{LU}_{-T:-1}$ of the LU, as well as the past evolution of the channel gain $g^{UU}_{-T:-1}$ of the UU. The future sum-rate of the UU is estimated based on a forecast $\hat{g}^{UU}_{0:\tau-1}$ of the evolution of the UU channel gain as
\begin{align}
     R^{UU}(P_{0:\tau-1},\hat{g}^{UU}_{0:\tau-1})=\frac{1}{\tau}\sum^{\tau-1}_{t=0}B\log_2\left(1+\frac{\hat{g}_t^{UU}P_t}{N_0B}\right),
\end{align}
where we recall that $P_t$ is the power allocated by the BS at time $t$.

The interference constraint for the LU is formulated as an upper bound on the expected maximum cumulative interference over $k$ subsequent communication slots. For a power allocation $P_{0},\dots,P_{\tau-1}$, the \emph{maximum $k$-step cumulative interference} at the LU over the future time horizon $0,1,\dots,\tau$, is measured as
\begin{align}
\phi\left(g_{0:\tau-1}^{LU},P_{0:\tau-1}\right)=\max^{}_{t'\in\{0,\dots,\tau-k-1\}}\frac{1}{k}\sum^{t'+k}_{t=t'}g_t^{LU}P_t,
    \label{eq:max_cum_k_int}
\end{align}
where the maximization ranges over all possible periods of $k$ times slots.
Therefore, for a safety threshold $\gamma$, the LU interference constraint amounts to the inequality
\begin{align}
    \mathbb{E}\left[\phi\left(g_{0:\tau-1}^{LU},P_{0:\tau-1}\right)\right]\leq   \gamma,
    \label{eq:max_int_constraint}
\end{align}
where the expectation is over the unknown LU channel evolution $g_{0:\tau-1}^{LU}$. We note that the constraint (\ref{eq:max_int_constraint}) can be also expressed in the language of robust signal temporal logic (STL) \cite{deshmukh2017robust}.

The interference threshold $\gamma$ in (\ref{eq:max_int_constraint}) is determined based on the observed past LU channel evolution $g^{LU}_{-T:-1}$. Accordingly, it is set to a fraction $\beta\in[0,1]$ of the maximum $k$-step cumulative interference over the past $T$ communication slots, i.e.,
\begin{align}
    \gamma=\beta\left(\max^{}_{t'\in\{-T,\dots,-1-k\}}\frac{ P_{\textrm{max}}}{k}\sum^{t'+k}_{t=t'}g_t^{LU}\right)
    \label{eq:max_th}.
\end{align}
By (\ref{eq:max_th}), a smaller value of $\beta$ imposes a stricter interference constraint.
The power control problem can then be formalized as the following \emph{open-loop} problem
\begin{subequations}
\begin{align}
\maximize_{P_{0}\dots,P_{\tau-1}}  \quad & \frac{1}{\tau}\sum^{\tau-1}_{t=0}B\log_2\left(1+\frac{\hat{g}_t^{UU}P_t}{N_0B}\right)\\
\textrm{s.t.} \quad & 0\leq P_t\leq P_{\textrm{max}}, \quad \forall t\in\{0,\dots,\tau-1\}\\
& \mathbb{E}\left[\phi\left(g_{0:\tau-1}^{LU},P_{0:\tau-1}\right)\right]\leq   \gamma \label{pc_max_int_mpc_constraint} .
\end{align}
\label{pc_max_int}%
\end{subequations} 
The constraint (\ref{pc_max_int_mpc_constraint}) is not directly tractable due to the expectation over the unknown distribution of the future evolution of the LU channel gain $g_{0:\tau-1}^{LU}$. However, the constraint function satisfies Assumption \ref{ass:lipshitz} in Sec. \ref{sec:mpc} by virtue of the inequality
\begin{align}
    \left|\phi\left(g_{0:\tau-1}^{LU},P_{0:\tau-1}\right)-\phi\left(\tilde{g}_{0:\tau-1}^{LU},P_{0:\tau-1}\right) \right|
    &\leq \frac{P_{\textrm{max}}}{k}\max_{t'\in\{0,\tau-1-k\}}\sum^{t'+k}_{t=t'}\left|g_t^{LU}-\tilde{g}_t^{LU}\right|.
\end{align}
As stated in Theorem \ref{th:mpc_exp_open_loop}, it is then possible to replace constraint (\ref{pc_max_int_mpc_constraint}) with a stricter constraint that depends on the output of a reliable set predictor. Accordingly, the solution of the resulting MPC problem yields a communication rate that lower bounds the optimal rate of the original problem (\ref{pc_max_int}), which is not attainable due to the lack of knowledge of future channel realizations. We benchmark the performance of different set predictors by addressing the associated MPC control problem.

\begin{figure}[t]
     \centering
     \begin{subfigure}[b]{0.49\textwidth}
         \centering
         \includegraphics[width=\textwidth]{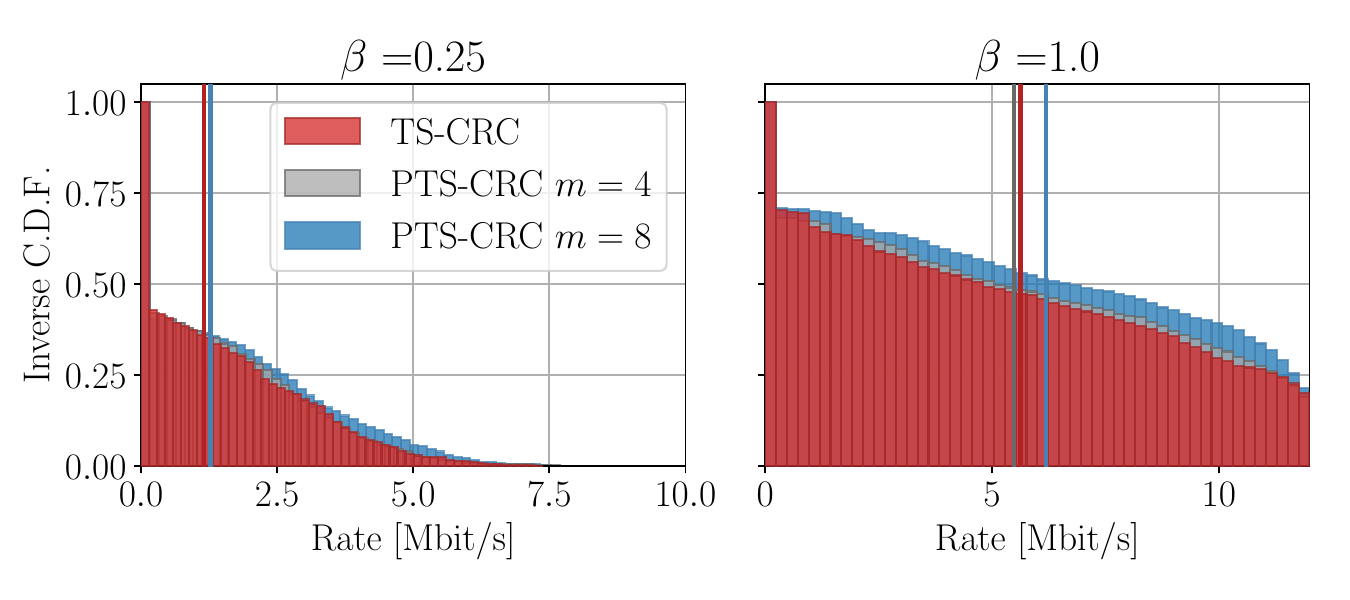}
         \caption{$k=1$\label{fig:beta01}}
     \end{subfigure}
     \hfill
     \begin{subfigure}[b]{0.49\textwidth}
         \centering
         \includegraphics[width=\textwidth]{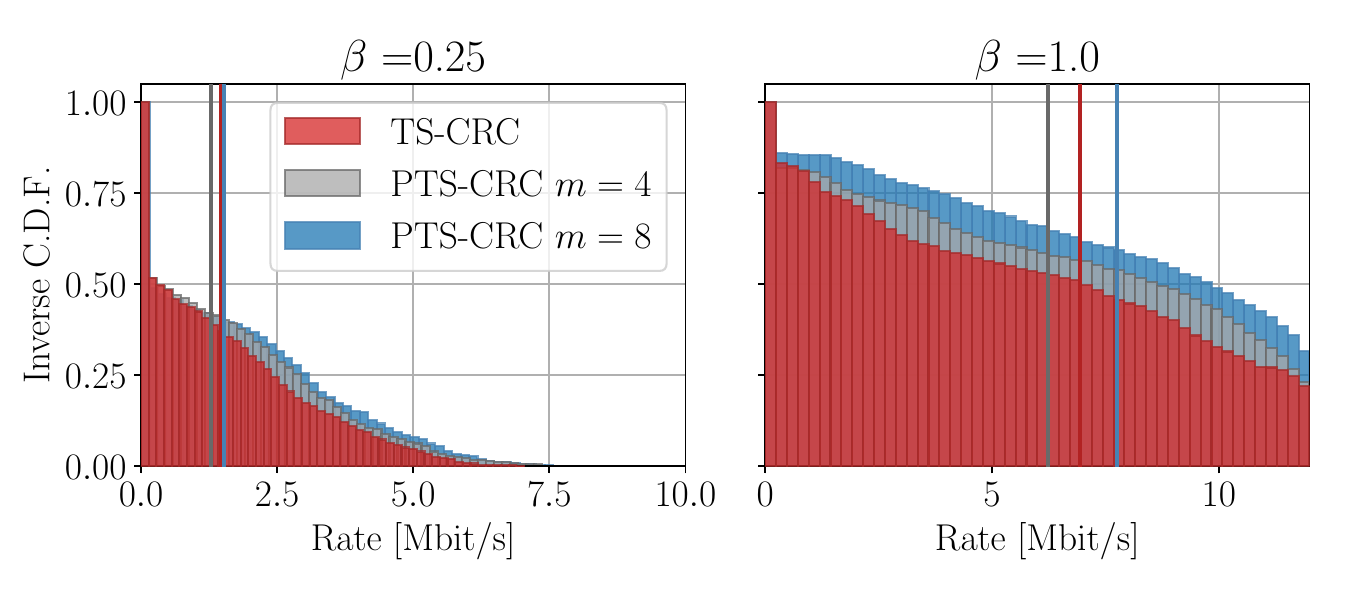}
         \caption{$k=3$\label{fig:beta025}}
     \end{subfigure}
    \caption{Inverse empirical cumulative distribution function (C.D.F.) of the rate obtained by the different model predictive power control policies as a function of the maximum LU interference level $\beta$.  For interference windows of size (a) $k=1$ and (b) $k=3$.}
       \label{fig:max_int_mpc} 
\end{figure}

In Fig. \ref{fig:example_pc}, we illustrate the power control solutions obtained by the proposed PST-CRC predictor with $m=8$ prototypes, and by a TS-CRC predictor, which is obtained by applying the same steps as PTS-CRC with $m=1$ to the predictive mean of the DeepAR model. This approach is adopted here as a benchmark since TS-CP \cite{stankeviciute2021conformal} cannot address the average constraint in (\ref{pc_max_int}). We set $T=30$ and $\tau=6$. While both power allocations meet the interference requirement, the PTS-CRC power control policy has larger transmit power, and therefore it attains a higher communication rate. This improvement is attributed to the higher efficiency of the PTS-CRC predictor, which leads to surrogate constraints that are less conservative compared to those given by TS-CRC.

The performance gain of PTS-CRC is further validated in Fig. \ref{fig:max_int_mpc}, in which we provide the inverse empirical cumulative distribution function (C.D.F) of the communication rate obtained by solving 1000 instances of the surrogate control problem for $\beta\in\{0.25,1\}$ and $k\in\{1,3\}$. As $\beta$ and $k$ increase, the interference constraint (\ref{pc_max_int_mpc_constraint}) is relaxed, and all power control policies yield larger communication rates. However, for fixed values of $\beta$ and $k$, the empirical C.D.F. of the PST-CRC power control policy has a heavier tail and a larger mean, indicating that these power allocations are able to serve the UU with larger rates. The performance gain becomes more evident for larger values of the number of predictor's samples $m$. For example, for $k=3$ and $\beta=1$, the 50th percentile of the PTS-CRC-based power control policy with $m=8$ prototypes is 80\% larger as compared to TS-CRC.

\begin{figure}[htp]
     \centering
     \begin{subfigure}[b]{0.24\textwidth}
         \centering
         \includegraphics[width=1\textwidth]{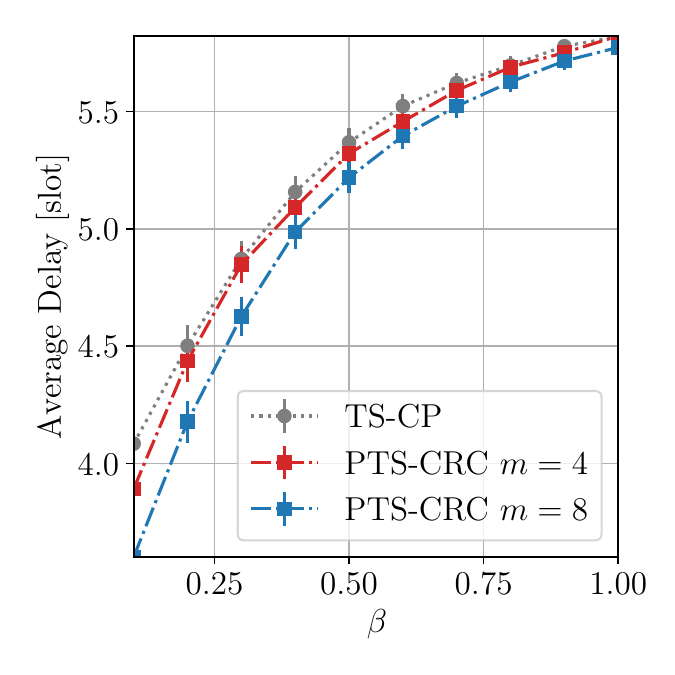}
         \caption{Average Delay\label{fig:delay}}
     \end{subfigure}
     \hfill
     \begin{subfigure}[b]{0.24\textwidth}
         \centering
         \includegraphics[width=1\textwidth]{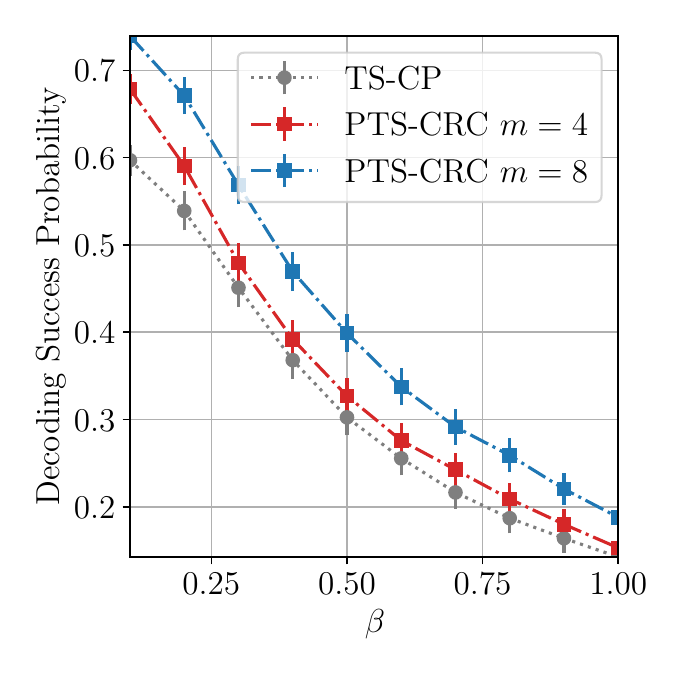}
         \caption{Decoding Probability \label{fig:dec_prob}}
     \end{subfigure}
     \hfill
     \begin{subfigure}[b]{0.24\textwidth}
         \centering
         \includegraphics[width=1\textwidth]{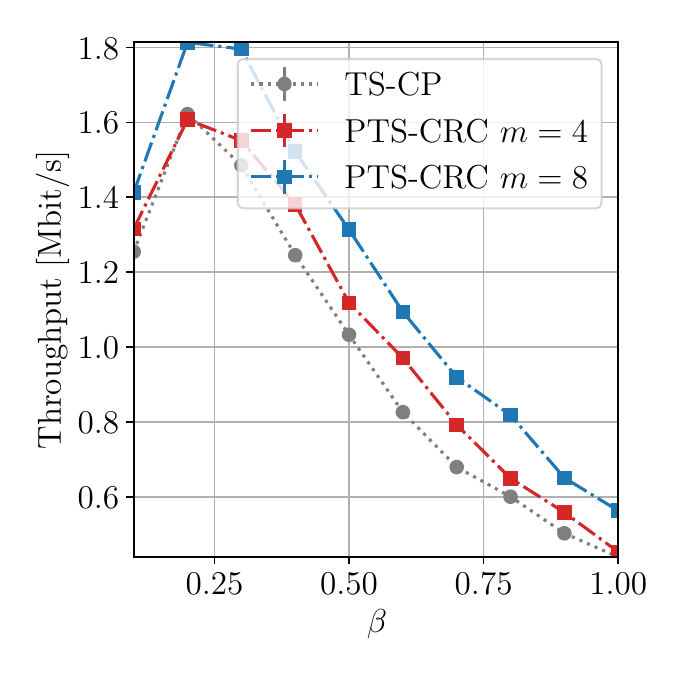}
         \caption{Throughput\label{fig:throughput}}
     \end{subfigure}
      \hfill
     \begin{subfigure}[b]{0.24\textwidth}
         \centering
         \includegraphics[width=1\textwidth]{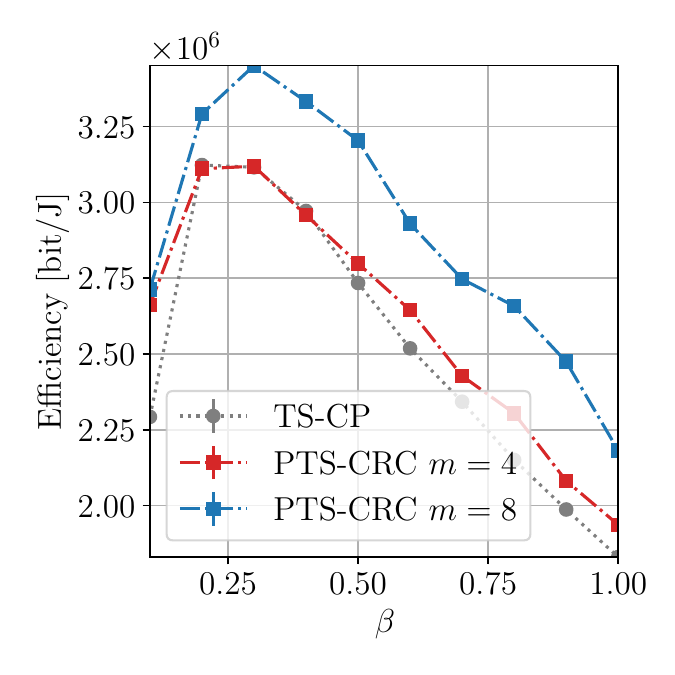}
         \caption{Energy Efficiency \label{fig:energy_eff}}
     \end{subfigure}
    \caption{(a) Average delay, (b) decoding probability, (c) throughput and (c) energy efficiency of HARQ-IR schemes based on the TS-CP predictor \cite{stankeviciute2021conformal} and the proposed PTS-CRC predictor for $m=4$ and $m=8$ prototypes.}
       \label{fig:predictionlength6} 
\end{figure}
\subsection{Energy Efficient HARQ-IR via Closed-Loop Model Predictive Power Control}

In this subsection, we address the problem of designing energy-efficient hybrid automatic repeat request with incremental redundancy (HARQ-IR) protocols \cite{lin2004error} by leveraging reliable channel state information forecasting. 

As a brief review, given a sequence of random channel gains $g_{t:\tau-1}$	and transmit powers $P_{t:\tau-1}$, $\tau-1-t$ retransmissions of a  packet encoded with rate $R$ [bit/s/Hz] yields successful decoding at the receiver with probability \cite{lee2015harq}
\begin{align}
	P_{dec}(P_{t:\tau-1},R)=\Pr[\sum^{\tau-1}_{t'=t}\log_2\left(1+\frac{P_{t'}g_{t'}}{BN_0}\right)> R],
	\label{eq:decoding_prob}
\end{align}
where the probability is over the future evolution of the channel gain $g_{t:\tau-1}$. 

At each time $t$, the base station (BS) has access to the feedback sequence $g_{-T:t-1}$ of past channel gains fed back by a user equipment and it must modulate the transmit power $P_t$ for the current time slot. The goal is to minimize energy expenditure while ensuring a minimum HARQ-IR decoding probability.
The target communication rate $R$ in constraint (\ref{eq:decoding_prob}) is set to the rate achieved during the $\tau$ communication slots $-\tau-1,\dots,-1$ prior to their start of the HARQ process for a  transmit power $\beta P_{\textrm{max}}$ with $\beta\in\{0,1\}$, i.e.,
\begin{align}
    R=\sum^{-1}_{t=-\tau-1}\log_2\left(1+\frac{\beta P_{\textrm{max}}g_t}{BN_0}\right).
    \label{eq:targer_rate}
\end{align}
By (\ref{eq:targer_rate}), a larger value of $\beta$ indicates a more stringent constraint  (\ref{eq:decoding_prob}).

Overall, at every time step $t=0,1,\dots,\tau-1$ the BS optimizes the transmit power level $P_t$ by addressing the \emph{closed-loop} MPC problem

\begin{subequations}
\begin{align}
\minimize_{P_{t}\dots,P_{\tau-1}}  \quad & \sum^{\tau-1}_{k=t}P_k\\
\textrm{s.t.} \quad & t\leq P_k\leq P_{\textrm{max}}, \quad \forall k\in\{t,\dots,\tau-1\}\\
& P_{dec}(P_{t:\tau-1},R-R_t)>\delta \label{HARQ_mpc_constraint} .
\end{align}
\label{HARQ_mpc}%
\end{subequations} 
where $0<\delta<1$ is the target decoding probability and 
\begin{align}
	R_t=\sum^{t-1}_{t'=0}\log_2\left(1+\frac{P_{t'}g_{t'}}{BN_0}\right)
\end{align}
 is the achieved rate decodable based on the past $t-1$ retransmissions. If the problem (\ref{HARQ_mpc}) is not feasible, the BS does not transmit, while if it is feasible the power $P_t$ is used for transmission. If $R-R_t<0$ the HARQ-IR message is successfully decoded, and hence the transmission process stops. The protocol also stops at the maximum number of retransmissions $\tau$ irrespective of whether the decoding was successful or not.

Constraint (\ref{HARQ_mpc_constraint}) cannot be evaluated, since the distribution of the future channel gain sequence $g_{t:\tau-1}$ is unknown. However, the constraint function satisfies Assumption \ref{ass:lipshitz} in Section \ref{sec:mpc}, since the inequality 	
\begin{align}
    \Bigg|\frac{1}{\tau}\sum^{\tau-1}_{t=0}\log_2\left(1+\frac{P_{t}g_t}{BN_0}\right)-\frac{1}{\tau}\sum^{\tau-1}_{t=0}\log_2\left(1+\frac{P_{t}\tilde{g}_t}{BN_0}\right)\Bigg|\leq \frac{ P_{\textrm{max}}}{B N_0}\frac{1}{\tau}\sum^{\tau-1}_{t=0}|{g}_t-\tilde{g}_t|.
\end{align}
Thus, by Theorem \ref{th:mpc_exp_closed_loop}, it is possible to replace the original constraint (\ref{HARQ_mpc_constraint}) with a stricter constraint based on the output of reliable set predictors. By solving the resulting optimization problem, we can obtain a power control policy that satisfies the reliability constraint.

For a numerical example, we set the observed feedback sequence of length $T=30$ and a maximum number of retransmissions $\tau=6$ steps using the TS-CP and the proposed TS-CRC set predictor with $m=4$ and $m=8$ prototypes. In Fig. \ref{fig:predictionlength6}, we present key performance indicators of the resulting HARQ-IR transmission protocol for different values of $\beta$. Specifically, we solve 1000 problem instances and compute the average number of retransmissions, the probability of decoding the HARQ packet, the average throughput,  and the average energy efficiency expressed as the number of decoded bits per Joule of transmit energy. 

 As the value of $\beta$ increases, the target information rate becomes larger, resulting in an increase in average delay and a decrease in decoding success probability for all schemes. However, when the PTS-CRC predictor is employed to predict the evolution of the future channel gain, the resulting HARQ-IR protocol can decode a larger fraction of information packets with shorter delays as compared to the TS-CP-based HARQ-IR scheme. Consequently, the PTS-CRC-based HARQ-IR scheme achieves an average throughput and average energy efficiency up to 25\% higher than that of the TS-CP-based scheme.

\section{Conclusions}\label{sec:conclusions}
In this work, we have addressed the problem of monitoring and controlling cyber-physical systems based on  set predictors that provide reliable uncertainty estimates. To this end, we have proposed PTS-CRC, a novel post-hoc calibration technique that leverages pre-trained probabilistic sequence models, like language models, to obtain predictive intervals with finite-sample reliability guarantees. PTS-CRC leverages an ensemble of prototype trajectories sampled from the sequence model  to effectively capture forking uncertainties,  while satisfying reliability guarantees beyond the conventional coverage criterion. Furthermore, we have demonstrated an application of PTS-CRC to open-loop and closed-loop model predictive control problems under general average constraints on the quality or safety of the control policy.

This paper has focused on settings in which the predictor or controller has access to calibration data in the form of sample sequences for the quantity to be predicted or for the target process. In an alternative setting, the predictor or controller may receive feedback on its predictions or actions in an online fashion without having offline access to calibration data. This setup was studied in \cite{gibbs2021adaptive,zaffran2022adaptive} for prediction and \cite{lekeufack2023conformal} for control. Integrating the methods proposed in this paper, which can address forking uncertainties, within the online setting is an interesting direction for future work.

\bibliography{biblio} 
\bibliographystyle{ieeetr}

  \appendices
\newpage
    \setcounter{page}{1}
  \section{Proofs}
     
    \subsection{Proof of Theorem \ref{th:reliability_implicit}}
    \label{sec:proof_th}
The proof of Theorem~\ref{th:reliability_implicit} relies on an integration of techniques introduced for the analysis of PCP \cite{wang2022probabilistic} and CRC \cite{angelopoulos2022conformal} (see Section \ref{sec:derivation}). The key idea is to treat the pair $(y_{-T:\tau-1}, \mathcal{P}^m)$ of true sequence $y_{-T:\tau-1}$ and prototype set $\mathcal{P}^m$, as a data point. The corresponding $n$ augmented calibration data points and the augmented test data point remain i.i.d., or more generally exchangeable, since the predicted sequences are conditionally independent given the respective past samples $\{y^i_{-T:\tau-1}\}_{i=1}^n$ and $y_{-T:\tau-1}$.
    
To elaborate, we simplify the notation by denoting the loss $\mathcal{L}(\Gamma_\lambda( y_{0:\tau-1}^i))$ of $i$-th calibration data point as $l_\lambda^i$, the test point as $y_{-T:\tau-1}^{n+1}$, as the test data, and the corresponding test loss as $l_\lambda^{n+1}$. We can then rewrite the expected test reliability requirement of PTS-CRC as $\mathbb{E}\big[l_{\lambda^{PTS-CRC}}^{n+1}\big] \leq \alpha,$
    where the expectation is taken as explained in Theorem \ref{theorem1} . We now introduce a genie-aided threshold $\lambda^*$ that is based on an empirical estimate of the average loss (\ref{eq:reliabilityguarantee}) using a data set that incorporates both calibration and test data as
    \begin{align}
    \lambda^{*}:=\inf\left\{\lambda: \frac{1}{n+1}\sum^{n+1}_{i=1}l^i_\lambda\leq \alpha \right\}.
    \label{eq:lambda_genie_aided}
    \end{align}
    Given Assumption~\ref{ass:loss}, we have the inequality $\lambda^{PTS-CRC}\geq \lambda^*$, for the threshold (\ref{right_lambda}) selected by PTS-CRC. since $\lambda^{PTS-CRC}$. Denote the (unordered) set of data points as $\Lbag y_{-T:\tau-1}^{1},...,y_{-T:\tau-1}^{n+1} \Rbag$. Using the law of total expectation, and given the monotonicity of the loss (Assumption \ref{ass:loss}), the average loss (\ref{eq:reliabilityguarantee}) can be bounded as
    \begin{align} 
        \mathbb{E}[l^{n+1}_{\lambda^{PTS-CRC}}] &\leq \mathbb{E}[l^{n+1}_{\lambda^{*}}] \\&= 
        \mathbb{E}\big[\mathbb{E}\big[l^{n+1}_{\lambda^{*}} | \Lbag l_{\lambda^*}^{1},...,l_{\lambda^*}^{n+1} \Rbag\big]\big]
        \nonumber\\&= 
        \mathbb{E}\bigg[ \frac{\sum_{i=1}^{n+1} l_{\lambda^*}^i}{n+1} \bigg] \leq \alpha,
    \end{align}
    where the second equality follows from the exchangeability of the losses $\{l^1_{\lambda^*},\dots,l^{n+1}_{\lambda^*}\}$. This concludes the proof.
  \subsection{Proof of Theorem \ref{th:mpc_exp_open_loop}}
\label{app:proof_mpc_exp_open_loop}
    By Assumption \ref{ass:loss},  from Theorem \ref{th:reliability_implicit}, the calibrated set predictor $\Gamma^{PTS-CRC}$ satisfies the inequality
\begin{align}
    \mathbb{E}\left[ \mathcal{L}\left(\Gamma^{PTS-CRC}(y_{-T:-1}),y_{0:\tau-1}\right)\right]= \mathbb{E}\left[\min_{\tilde{y}_{0:\tau-1}\in\Gamma^{PTS-CRC}(y_{-T:-1})} m(y_{0:\tau-1},\tilde{y}_{0:\tau-1})\right]\leq \alpha.
    \label{eq:inequalitybetareliable}
\end{align} 
From Assumption \ref{ass:lipshitz}, for any pair $(y_{0:\tau-1},s_{0:\tau-1})\in \mathcal{Y}^\tau\times \mathcal{S}^\tau$ and for any sequence $\tilde{y}_{0:\tau-1}\in \Gamma^{PTS-CRC}(y_{-T:-1})$, we have the inequality
\begin{align}
    c(s_{0:\tau-1},y_{0:\tau-1})&\leq  c(s_{0:\tau-1},\tilde{y}_{0:\tau-1})+ L m(y_{0:\tau-1},\tilde{y}_{0:\tau-1}).
    \label{eq:control}
\end{align}
Minimizing the right-hand size of (\ref{eq:control}) over $\tilde{y}_{0:\tau-1}$, we get
\begin{align}
c(s_{0:\tau-1},y_{0:\tau-1})&\leq  \min_{\tilde{y}_{0:\tau-1}\in \Gamma^{PTS-CRC}(y_{-T:-1})}\left[c(s_{0:\tau-1},\tilde{y}_{0:\tau-1})+ L m(y_{0:\tau-1},\tilde{y}_{0:\tau-1})\right]\\
&\leq  \max_{\tilde{y}_{0:\tau-1}\in \Gamma^{PTS-CRC}(y_{-T:-1})}c(s_{0:\tau-1},\tilde{y}_{0:\tau-1})+ L \min_{\tilde{y}_{0:\tau-1}\in \Gamma^{PTS-CRC}(y_{-T:-1})}m(y_{0:\tau-1},\tilde{y}_{0:\tau-1}).
\end{align}
Taking the expectation with respect to the test sequence $y_{0:\tau-1}\sim p(y_{0:\tau-1}|y_{-T:-1})$, we obtain
\begin{align}
\mathbb{E}\left[c(s_{0:\tau-1},y_{0:\tau-1})\right]\leq  \max_{\tilde{y}_{0:\tau-1}\in\Gamma^{PTS-CRC}(y_{-T:-1})}c(s_{0:\tau-1},\tilde{y}_{0:\tau-1})+ L \alpha, \label{eq:last_eq}
\end{align}
where the inequality follows from (\ref{eq:inequalitybetareliable}). We conclude that if there exists a sequence $u_{T+1}\dots,u_{T+\tau}$ such that associated state sequence $s_{T+1}\dots,s_{T+\tau}$ satisfies (\ref{eq:surrogate_mpc_constraint}), then $u_{T+1}\dots,u_{T+\tau}$ is in the feasible set of (\ref{mpc}) and it satisfies the desired inequality $
\mathbb{E}\left[c(s_{0:\tau-1},y_{0:\tau-1})\right]\leq 0.$

\subsection{Proof of Theorem \ref{th:mpc_exp_closed_loop}}
\label{app:mpc_exp_closed_loop}
Following the proof for the open-loop case (Theorem \ref{th:mpc_exp_open_loop}), if for every time step $t=0,\dots,\tau-1$ there exists a control sequence $u_{t:\tau-1}$ for which the control problem  is feasible, then the reliability guarantee $\mathbb{E}\left[c(s_{t:\tau-1},y_{t:\tau-1})|s_{0:t-1},y_{0:t-1}\right]\leq 0 $ 
is satisfied for all $t\in[0,\tau-1]$ and, therefore the original guarantee (\ref{mpc_constraint}) is satisfied.

\end{document}